\documentclass{aa}  

\usepackage{graphicx,txfonts,natbib,here,url,color,amsmath,fixltx2e}

\bibpunct{(}{)}{;}{a}{}{,}

\begin{document}

   \title{The association of the Hale Sector Boundary \\with RHESSI solar flares and active longitudes}

   \author{K. Loumou\inst{1} \and I. G. Hannah\inst{1} \and H. S. Hudson\inst{1,2}}

   \institute{SUPA School of Physics and Astronomy, University of Glasgow, Glasgow, G12 8QQ, UK\\ \email{k.loumou.1@research.gla.ac.uk}\\ \email{iain.hannah@glasgow.ac.uk}
         \and Space Sciences Laboratory, University of California, Berkeley, CA 94720, USA\\
            \email{hhudson@ssl.berkeley.edu}}

   \date{Received ; accepted }
   
   \titlerunning{Hale Sector Boundary, RHESSI flares and active longitudes}

  \abstract
   {The heliospheric magnetic field (HMF) is structured into large sectors of positive and negative polarity. The parts of the boundary between these sectors where the change in polarity matches that of the leading-to-following sunspot polarity in that solar hemisphere, are called Hale Sector Boundaries (HSB).} 
   {We investigate the flare occurrence rate near HSBs  and the association between HSBs and active longitudes.}
   {Previous work determined the times HSBs were at solar central meridian, using the detection of the HMF sector boundary crossing at the Earth. In addition to this, we use a new approach which finds the HSB locations at all times by determining them from Potential Field Source Surface (PFSS) extrapolations of photospheric magnetograms. We use the RHESSI X-ray flare list for comparison to the HSB as it provides accurate flare locations over 14 years, from February 2002 to February 2016, covering both Cycles 23 and 24. For the active longitude positions we use previously published work based on sunspot observations.}
   {We find that the two methods of determining the HSB generally agree and that $41\%$ (Cycle 23) and $47\%$ (Cycle 24) of RHESSI flares occur within $30^\circ$ of the PFSS determined-HSB. The behaviour of the HSBs varies over the two Cycles studied, and as expected they swap in hemisphere as the Cycles change. The HSBs and active longitudes do overlap but not consistently. They often move at different rates relative to each other (and the Carrington solar rotation rate) and these vary over each Cycle. The HSBs provide a useful additional  activity indicator, particularly during periods when active longitudes are difficult to determine.}
{}

   \maketitle

\section{Introduction}\label{nsec:intro}

\begin{figure*}
\centering
 \includegraphics[width=0.7\linewidth]{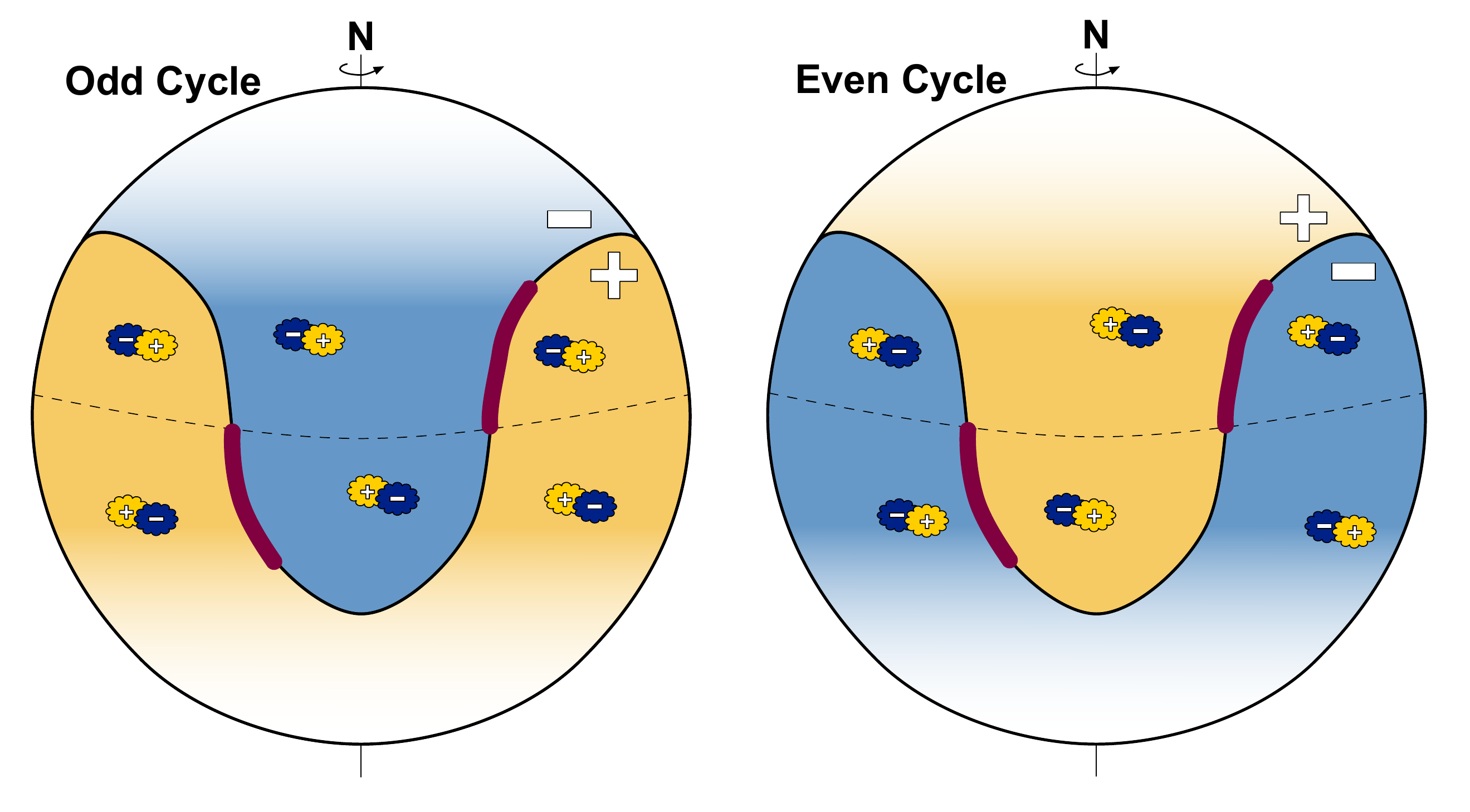}
\caption{Cartoon of the Hale Sector Boundary (thick purple lines) for odd (left) and even (right) Cycles during periods when four-sector boundary crossings would be detected at the Earth. The HSBs are the parts of the sector boundary where the magnetic polarity change, from leading-to-following (right to left in each panel), matches the magnetic polarity change of the sunspots in that hemisphere. Positive magnetic polarity is indicated by yellow, negative by blue. Based on the Cycle 20 example in \citet{1976SoPh...49..177S}.}
\label{fig:hsb_cartoon}
\end{figure*}

Predicting solar magnetic activity such as flares and coronal mass ejections is difficult, but high degrees of organisation in both time and position do exist. With each new 11-year activity Cycle, sunspots appear at high latitudes. As the cycle evolves, new groups appear closer to the equator, producing the classic Butterfly Diagram \citep{1904MNRAS..64..747M}. In addition, sunspots tend to occur in pairs with oppositely directed magnetic field and have the same configuration of leading-to-following polarity (in the direction of solar rotation) in each hemisphere. This is reversed between northern and southern hemispheres and then swaps for each solar Cycle, i.e. Hale's law \citep{1919ApJ....49..153H}. As the solar wind carries the magnetic field into the heliosphere, it simplifies into large sectors of different magnetic polarities separated by the Heliospheric Current Sheet (HCS) \citep{1973Ap&SS..24..371S}. This boundary separating the different sectors of magnetic polarity is detectable at the Earth. In this paper, we study its association with solar activity and also compare it with the phenomenon of active longitudes. 

 Concentration of activity in terms of longitude have been suggested since the time of \citet{1863spots...C}. Initial studies of ``active longitudes'' \citep[e.g.][]{1939POMic...7..127L} found substantial ambiguities in their existence and precise location. A copious literature has developed to characterise these regions, testing their usefulness for solar activity prediction and probe their source and implications for dynamo theory \citep[e.g.][]{2007AdSpR..40..951U}. Several terms have been used for these regions, including ``nests of activity'' \citep{1983ApJ...265.1056G,1987ARA&A..25...83Z}, but all these describe locations where there is a strong tendency for flux emergence to persist over long timescales. 
 
 The general approach for finding active longitudes is to filter the activity tracer, i.e. sunspots or flares, so that only the most dominant regions per time remain, often requiring a correction for rotation at rates differing from those of the Carrington synodic period \citep[e.g.][]{2007AdSpR..40..951U}. A variety of techniques have been applied to active longitude studies but these can sometimes produce artefacts \citep[e.g.][]{2010A&A...513A..48P} that can lead to ambiguities in the quantitative properties of active longitudes. The work of \citet{2003A&A...405.1121B} and \citet{2005A&A...441..347U} used the Greenwich sunspot data, finding two active longitudes in each hemisphere about $180^\circ$ apart. This pattern lasted for several solar Cycles. At any given time one of the active longitudes was more dominant than the other, with the activity ``flip-flopping'' between them. These active longitudes were affected by differential rotation but at a different rate to the sunspots and were asynchronous between hemispheres. \citet{2006A&A...445..703B} provided a qualitative explanation for their results in terms of the internal solar dynamo, with it either being a differentially rotating magnetic structure or a solid rotator with a stroboscopic effect producing the observed behaviour.  Similarly behaving active longitudes have been identified via EUV observations of coronal streamers \citep{2011ApJ...735..130L}, as well as flare locations during solar Cycles 19-23 \citep{2003ApJ...585.1114B} and Cycles 21-23 \citep{2011JASTP..73..258Z}.
 
  The Debrecen photoheliographic sunspot data have contributed to active longitude studies \citep{2012CEAB...36....9G,2014SoPh..289..579G,2016ApJ...818..127G}, leading to the conclusion that about 60\% of X-ray flares occurred with $\pm36^\circ$ of an active longitude. This work again shows two bands of activity, with one being more productive than the other and this varying over time (the ``flip-flop'' behaviour). These regions were narrow during the starting and decay phases of each Cycle (about $20^\circ-30^\circ$) but wider ($60^\circ$) during maximum. They also showed how the active longitudes' rotation rates vary over each Cycle. Initially they are faster than the Carrington rate but near solar maximum start to slow down, resulting in a ``parabolic migration path'' relative to the Carrington rate \citep{2016ApJ...818..127G}. This time evolution of their location was previously noted by \citet{2005A&A...441..347U}. The properties of the sunspot groups near active longitudes, such as complexity and helicity, were also found to be more preferential for CMEs to occur \citep{2017ApJ...838...18G}.
  
 Despite ambiguity about the specific nature of active longitudes it is clear that they have something to do with the internal dynamo of the Sun, and that this and other related phenomena should manifest themselves in the global solar magnetic field. The discovery of the sector boundaries in the heliospheric magnetic field \citep{1965JGR....70.5793W} provided an alternative physical framework for characterising the large-scale distribution of solar activity, and one with links to solar-terrestrial impacts via ``proton flares'' \citep{1969SoPh....6..104B}. We now understand the sector structure to be the result of the natural simplification of the Sun's large-scale magnetic field as the solar wind drags it out into an almost bipolar configuration \citep[e.g.][]{2013LRSP...10....5O}. The heliospheric current sheet (HCS) is the warped surface that separates the sectors of positive and negative polarity, and where this intersects the Earth's orbit we detect the polarity reversal. The HMF essentially has the polarity pattern of a dipole, but with an inverse-square falloff with radial distance as dictated by the solar wind. A simple tilted dipole, relative to the ecliptic, results in two sector boundary crossings detectable at the Earth. The seasonal variation of sector widths reveals the presence of the tilt \citep{1969JGR....74.5611R}. This two-sector structure is mostly seen during the weaker phases of the solar Cycle (declining and minimum) with a more complicated four-sector structure arising from the periods when the quadrupole component is relatively stronger compared to the dipole \citep[e.g.][]{2017SoPh..292..174G}. During these times, the HCS has a significant warp, as well as a tilt relative to the solar rotation axis.

Early work showed that the sector boundary locations, identified solely via the polarity reversal in the solar wind near 1~AU, was associated with enhancements in activity in terms of both the green line corona \citep{1974SoPh...36..115A} and flares \citep{1975SoPh...41..227D}. In the latter case, a superposed epoch analysis of flares from 1964 to 1970 (Cycle 20) showed a marked increase in occurrence with the negative leading positive (-,+) sector boundary crossing. This preference was more noticeable for northern hemisphere flares, where the magnetic polarity matched Hale's law. This pattern was further shown by \citet{1976SoPh...49..177S} for the green line corona, with the maximum occurring above the ``Hale Sector Boundary'' (HSB), the part of the sector boundary with the same polarity change as Hale's law gives for sunspots in that solar hemisphere. As the sunspots' leading magnetic polarities swap between alternate 11-year sunspot Cycles \citep{1919ApJ....49..153H}, the location of the HSB of a particular polarity change will be different for odd and even numbered Cycles. Fig.~\ref{fig:hsb_cartoon} illustrates this relationship for each Cycle during periods when four sectors are detected at the Earth. Specifically we have 

\begin{itemize}
  \item For an odd Cycle (left panel, Fig. \ref{fig:hsb_cartoon}), the positive leading negative $(+,-)$ HSB will be in the northern hemisphere while the negative leading positive $(-,+)$ will be in the southern hemisphere;
  \item For an even Cycle (right panel, Fig. \ref{fig:hsb_cartoon}), the positive leading negative $(+,-)$ HSB will be in the southern hemisphere while the negative leading positive $(-,+)$ will be in the northern hemisphere;
\end{itemize}

The sector boundaries detected via the crossing of the HCS at the Earth usually have clear signatures. From these epochs, a ballistic assumption about solar wind transport can be used to estimate how many days beforehand the structure was at central meridian at the Sun. \citet{2011ApJ...733...49S} used a range of 4.5 to 6.5 days, with superposed epoch analysis on magnetograms from Cycles 21 to 24 and X-ray flare positions, from RHESSI and GOES, covering 2002 to 2008 and 1996 to 2008 respectively. In both cases, there was a strengthening of activity in the expected hemisphere at central meridian corresponding to the HSB location. This approach was repeated for sunspots over Cycles 16 to 24, again finding a concentration at times when the HSB was expected to be at the central meridian in each hemisphere \citep{2014SSRv..186...17H}. It has also been shown that sunspot pairs are more likely to develop near the HSB \citep{2015GeoRL..42.2571A}. This work again showed that even though the structure is being detected at the Earth, it could be clearly related to a concentration of activity back at the Sun. \citet{2017SoPh..292..174G} carefully looked at the Earth sector boundary crossings for Cycles 21 to 24 compared to photospheric magnetograms. Instead of using a fixed timelag (constant solar wind speed) to get the HSB back at central meridian, they took into account the observed speeds, which produced a sharper mapping back to the Sun. Overall they found that this refined technique, compared to the approach of \citet{2011ApJ...733...49S}, yields similar patterns. With their refined HSB approach they found that the HSB mapping into the magnetograms was statistically significant, particularly during the maximum and declining phases of each Cycle. They also found that the clearest HSB association was in the northern hemisphere during the odd Cycles studied, but was weaker in the southern hemisphere during even Cycles. No difference was found between Cycles for the (-,+) HSB, with a consistent association with activity. 

In this paper we extend the work of \citep{2011ApJ...733...49S} by comparing the HSB found via the detection of the boundary crossing at the Earth (which we call HSB-Earth, to distinguish it from the HSB at the Sun) to RHESSI X-ray flares into Cycle 24, detailed in \S\ref{nsec:findhsbesc}. Then in \S\ref{nsec:hsbvsal}, we compare these HSBs to the active longitudes determined by \citet{2016ApJ...818..127G} over Cycles 21 to 24 to investigate how these phenomena relate. Even with the refined approach of \citet{2017SoPh..292..174G}, the HSB can only be found at times corresponding to central meridian passage. So to be able to determine the HSB locations for all times we instead determine it from daily Potential Field Source Surface (PFSS) maps, described in \S\ref{nsec:findhsbpfss} (which we call HSB PFSS). This also has the advantage that the HSB can be found closer to the Sun, minimising the uncertainty introduced by the variable solar wind speed in the HSB-Earth approach. We compare these two HSB approaches in \S\ref{nsec:escvspfss} and then determine the closest HSB PFSS for each RHESSI X-ray flare, investigating their association in \S\ref{nsec:pfssvsfl}.


\section{HSB-Earth and Flaring Rate}\label{nsec:findhsbesc}

\begin{figure*}[t]
\centering
\includegraphics[width=0.8\linewidth]{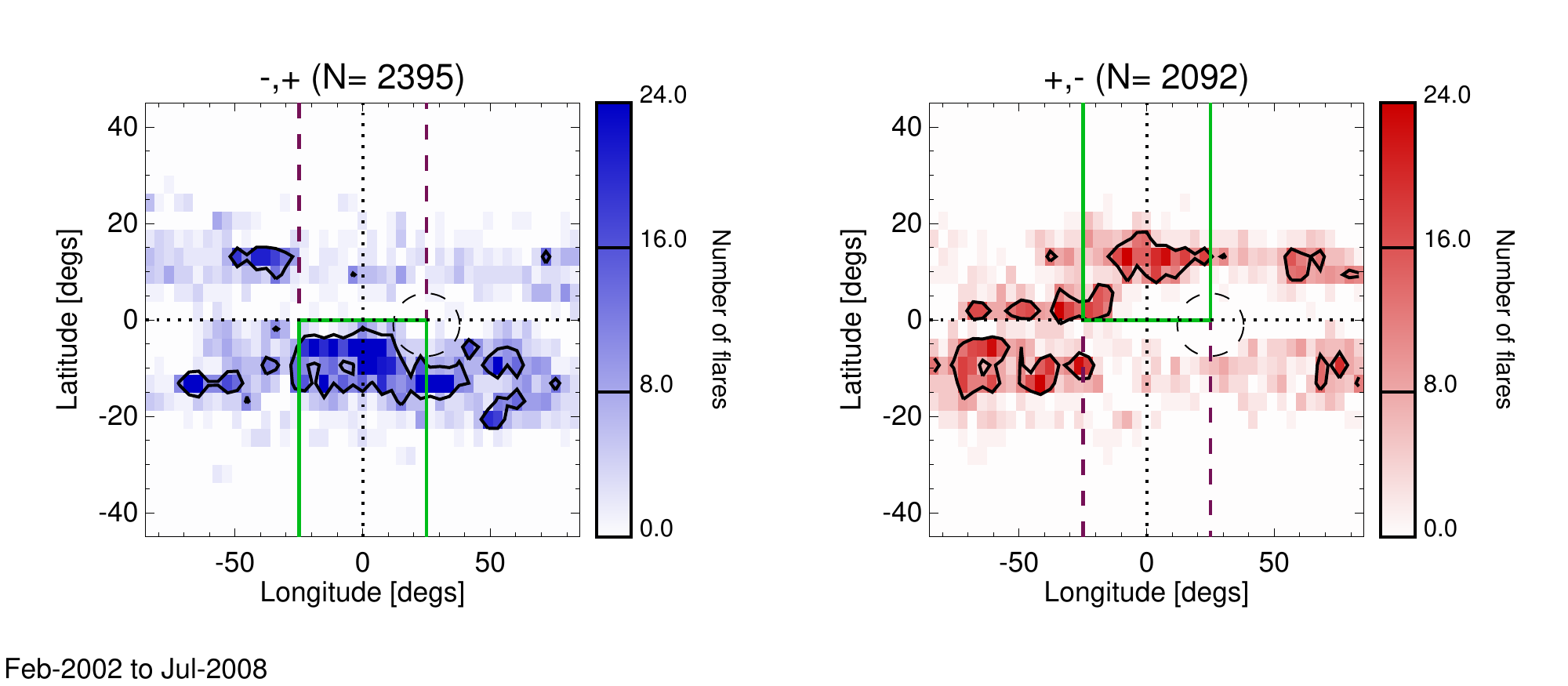}\\
\includegraphics[width=0.8\linewidth]{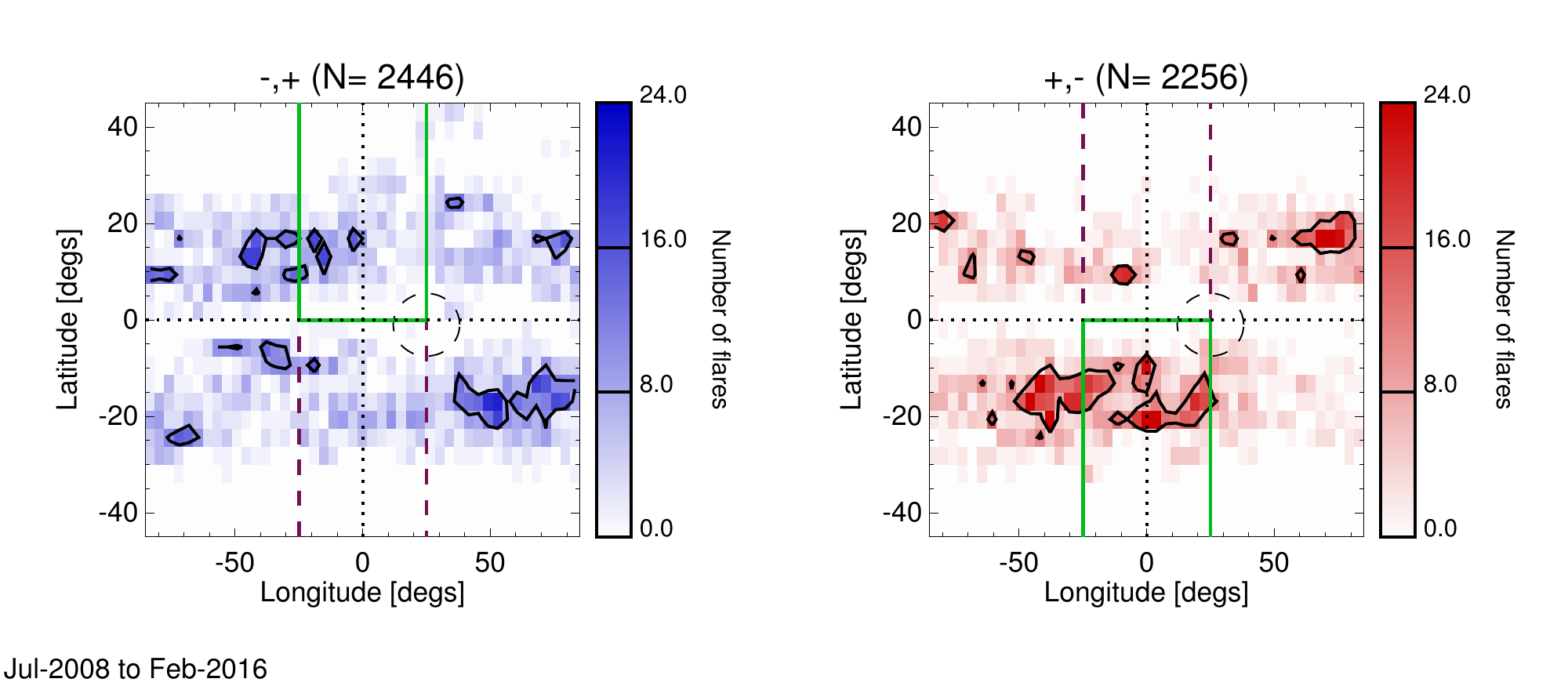}
\caption{Distribution of the RHESSI flares that occur when the HSB-Earth was at central meridian, separated by the polarity change at the boundary : left (blue) for negative leading positive (-,+), right (red) for positive leading negative (+,-). In the top row we show the flares during Cycle 23 (Feb 2002 to July 2008) and bottom row for Cycle 24 (July 2008 to Feb 2016). The green lines indicate the hemisphere in which the HSB is located. The dashed circle indicates the approximate region from which RHESSI cannot determine a flare's location. The purple dashed lines indicate $\pm 25^\circ$ of meridian, the range used for the histograms in Fig. \ref{fig:1D_all_flares}.}
\label{fig:2D_all_flares}
\end{figure*} 

\begin{figure*}[t]
\centering         
\includegraphics[width=0.4\linewidth]{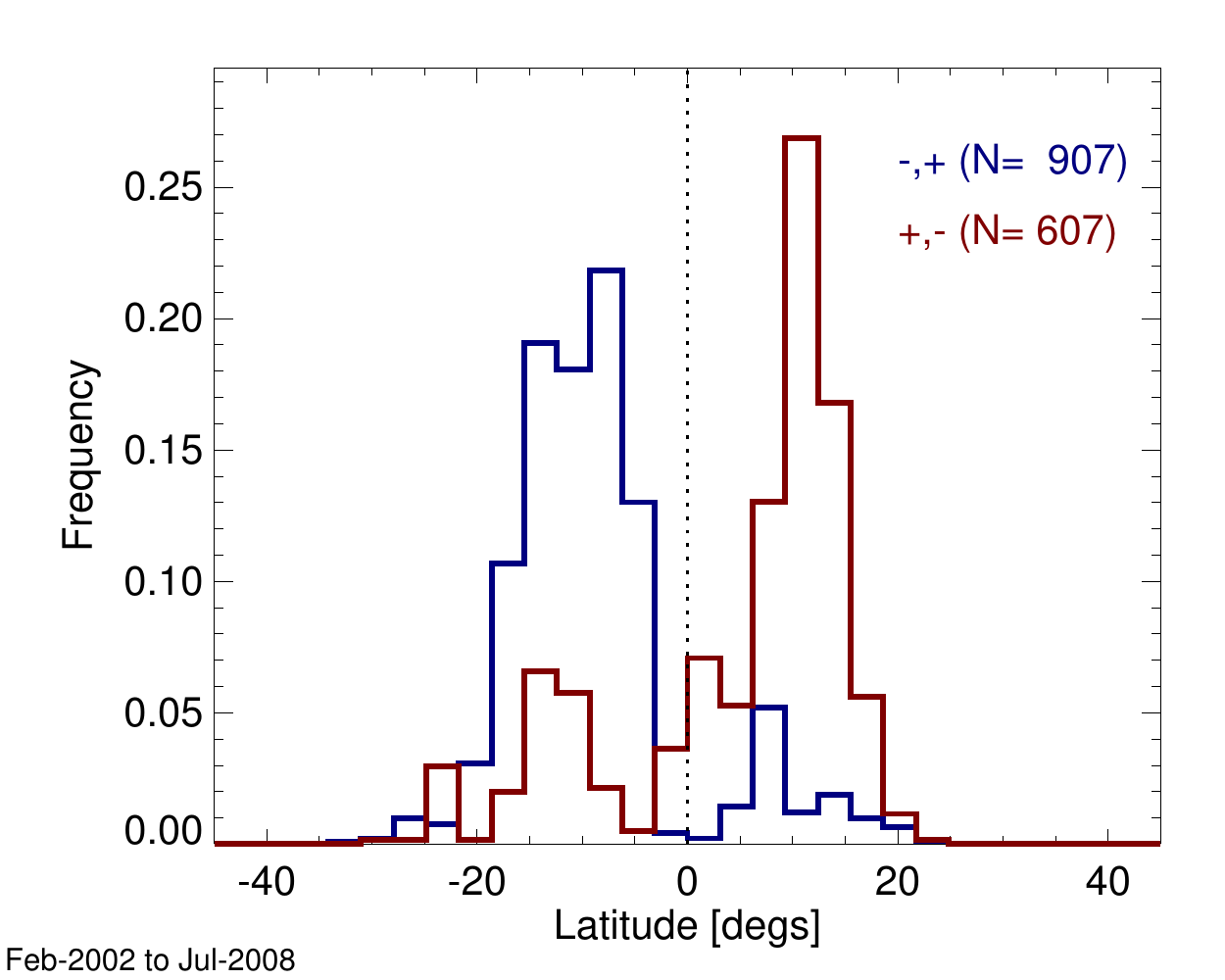}
\includegraphics[width=0.4\linewidth]{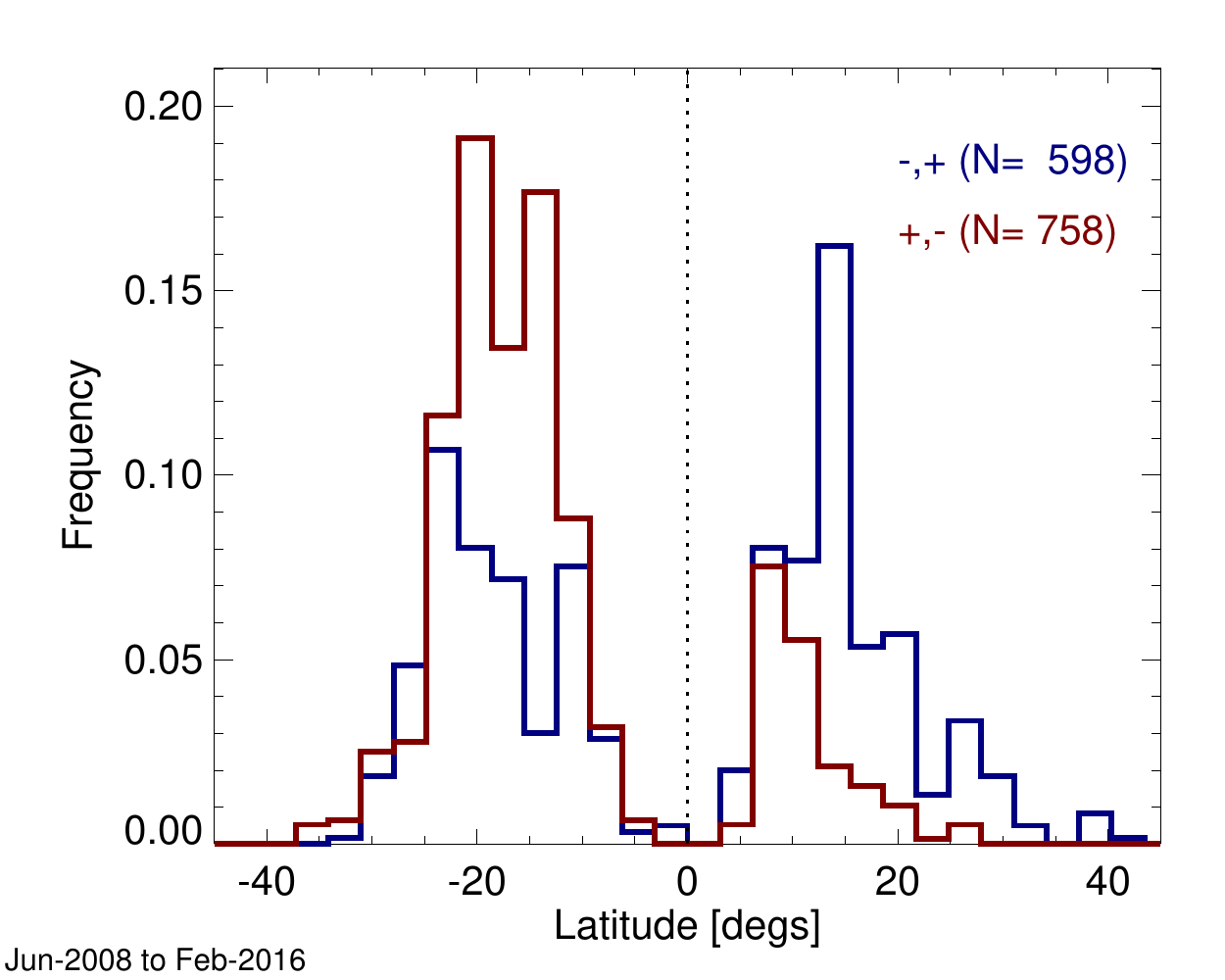}
\caption{Latitude histograms of the RHESSI flares with longitudes $|\lambda_{F}|\leqslant25^{\circ}$ during the times the HSB was at central meridian. The histograms are separated into those from Cycle 23 (left) and Cycle 24 (right) and by the polarity change at the sector boundary of positive leading negative ( +,- red) and negative leading positive (-,+ blue).}
\label{fig:1D_all_flares}
\end{figure*}  

 \begin{figure}
\centering         
\includegraphics[width=0.4\linewidth]{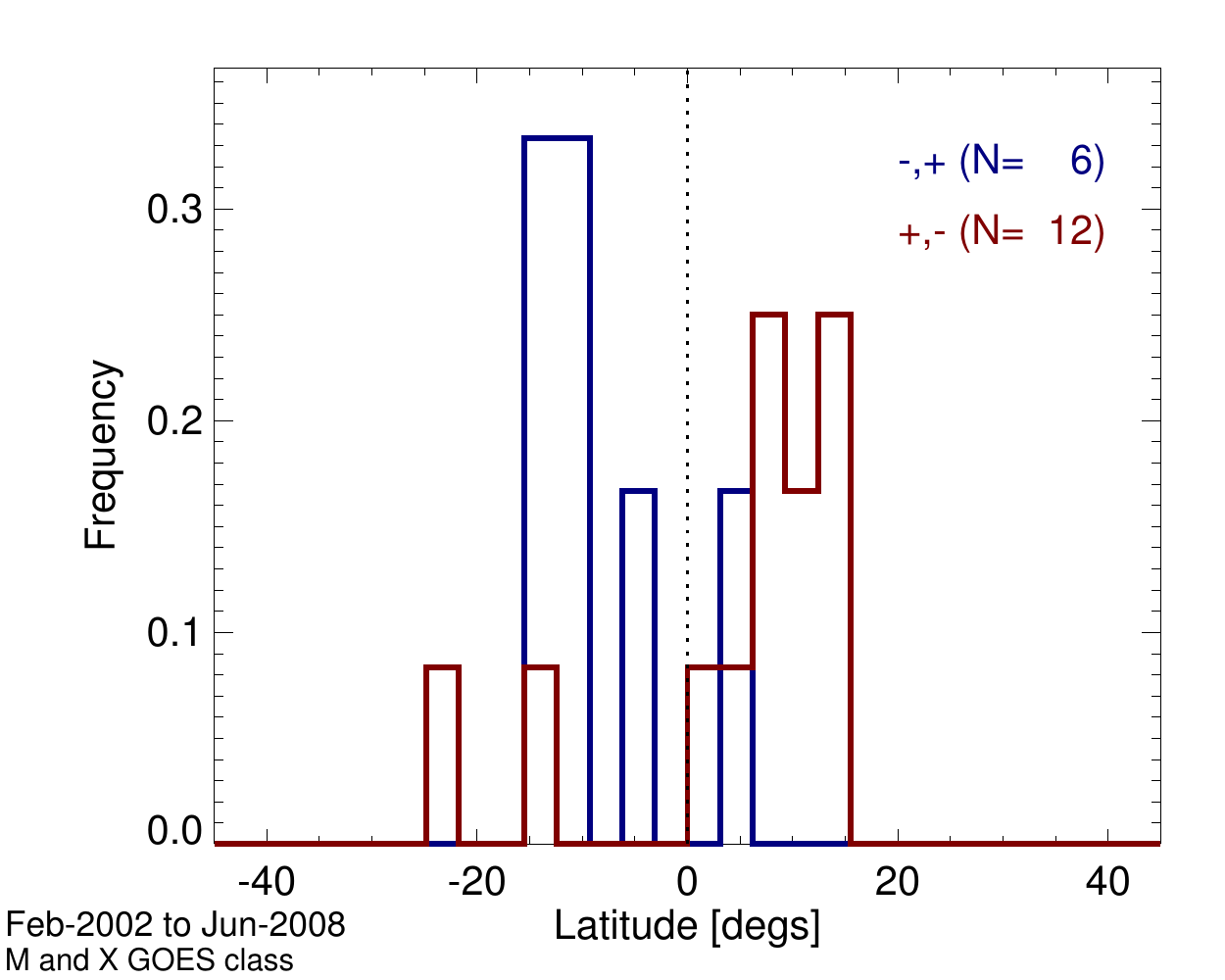}
\includegraphics[width=0.4\linewidth]{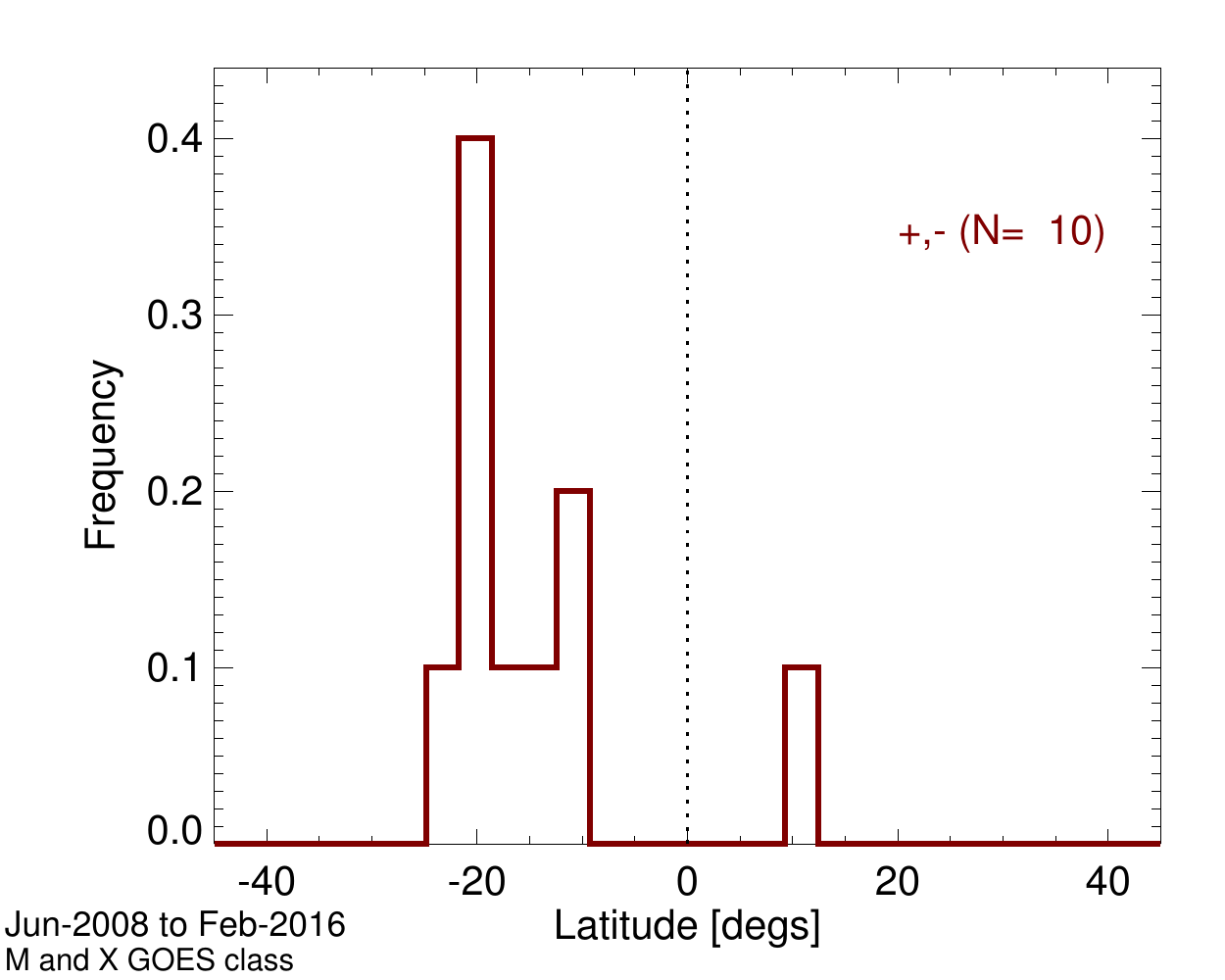}\\
\includegraphics[width=0.4\linewidth]{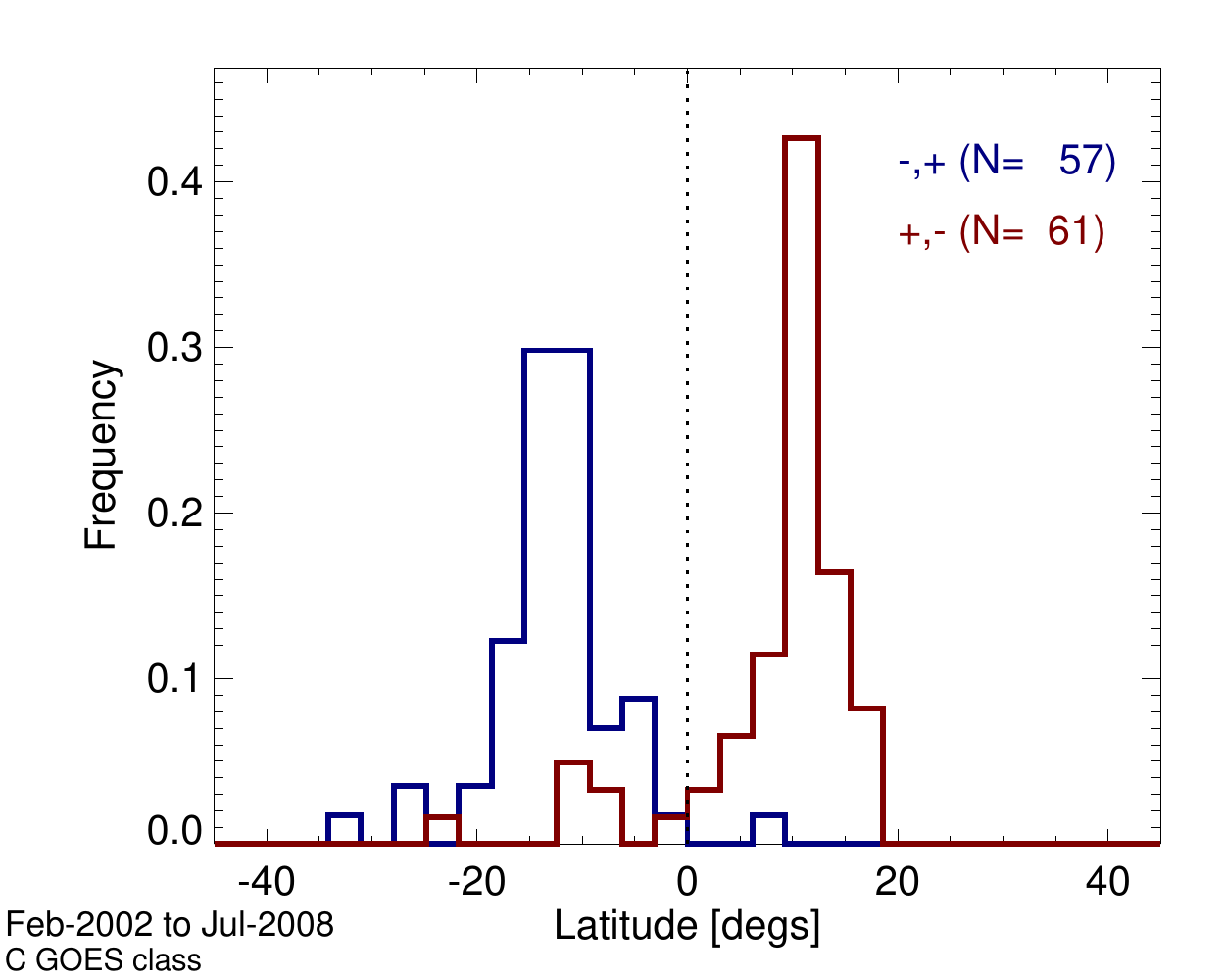}
\includegraphics[width=0.4\linewidth]{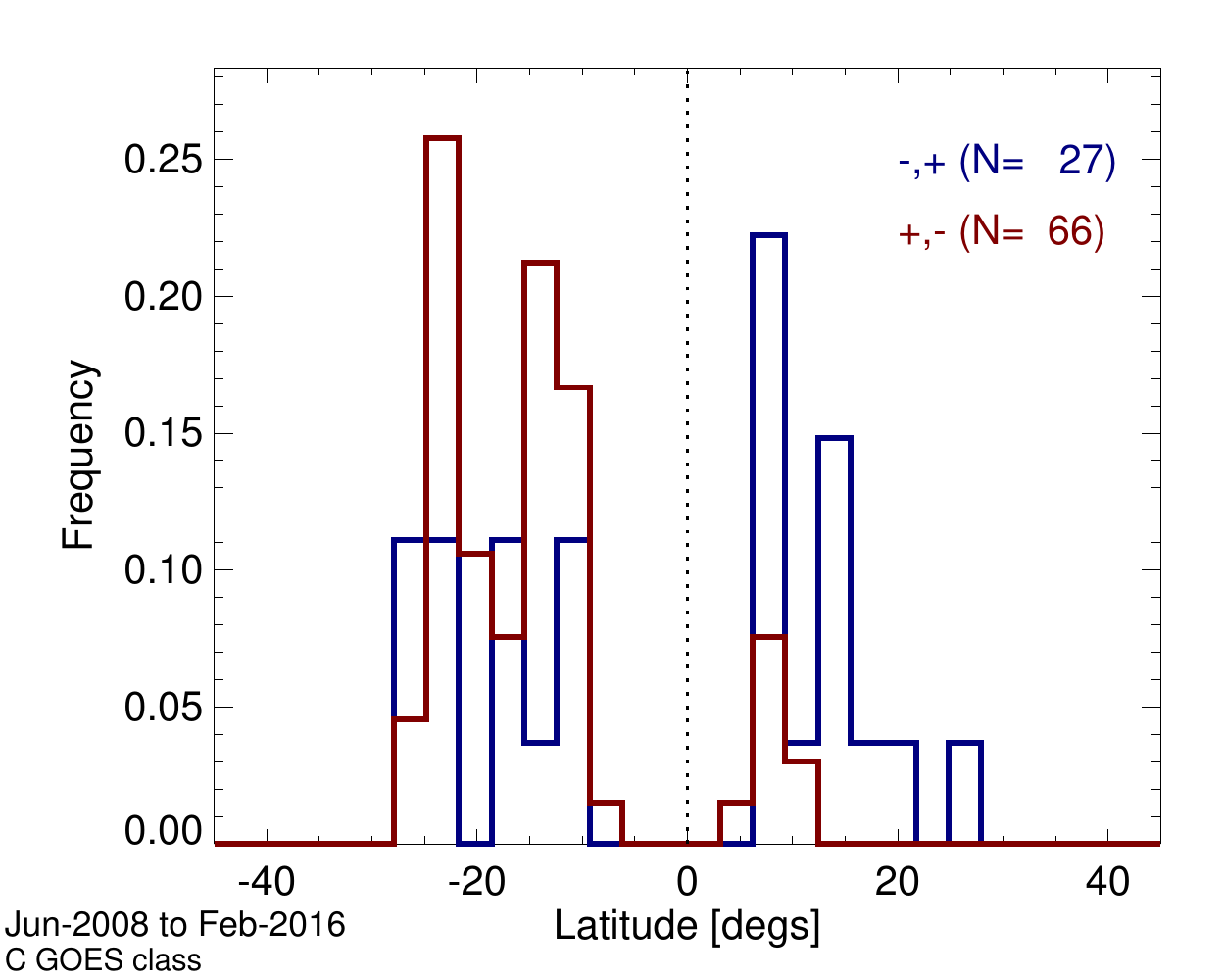}\\
\includegraphics[width=0.4\linewidth]{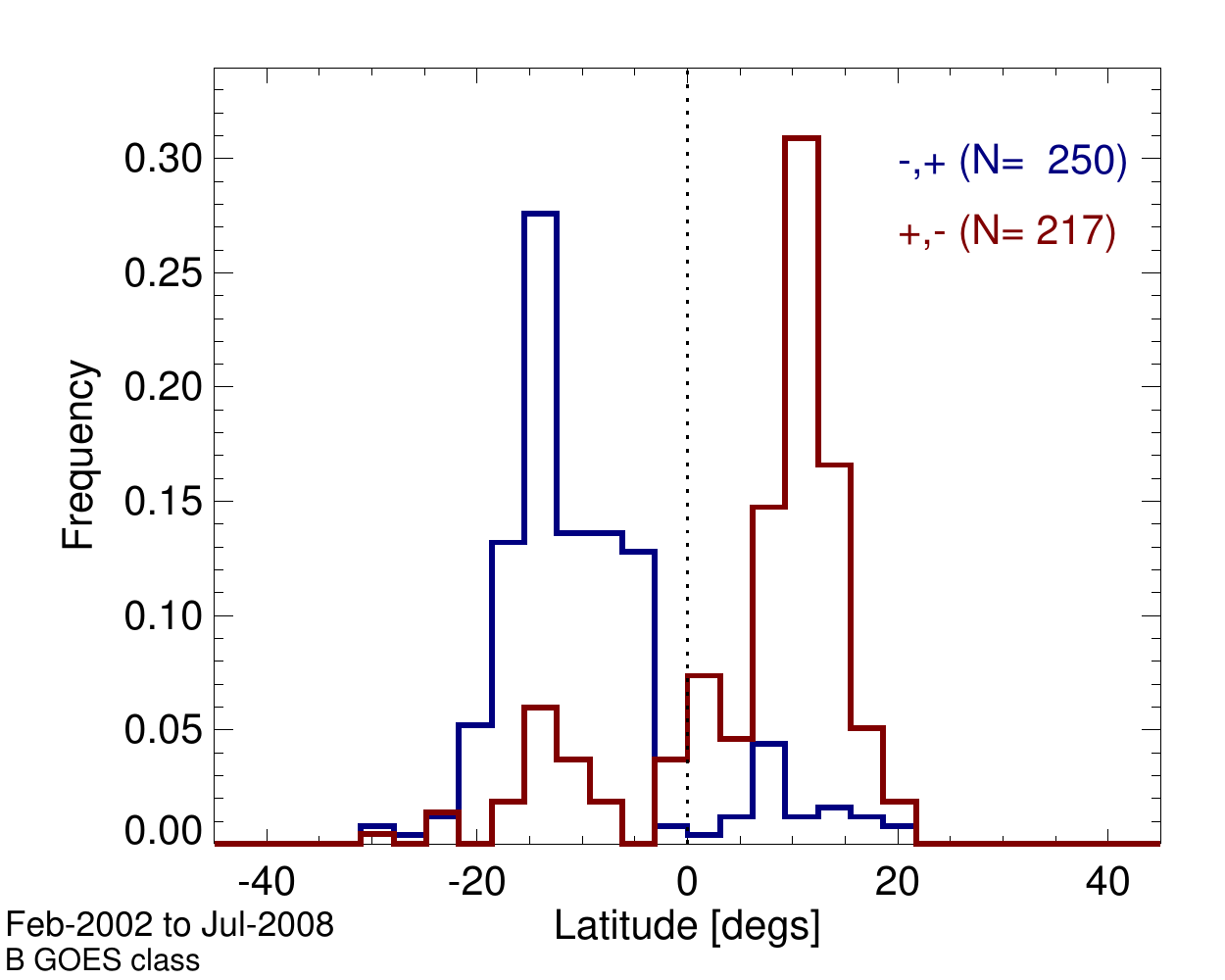}
\includegraphics[width=0.4\linewidth]{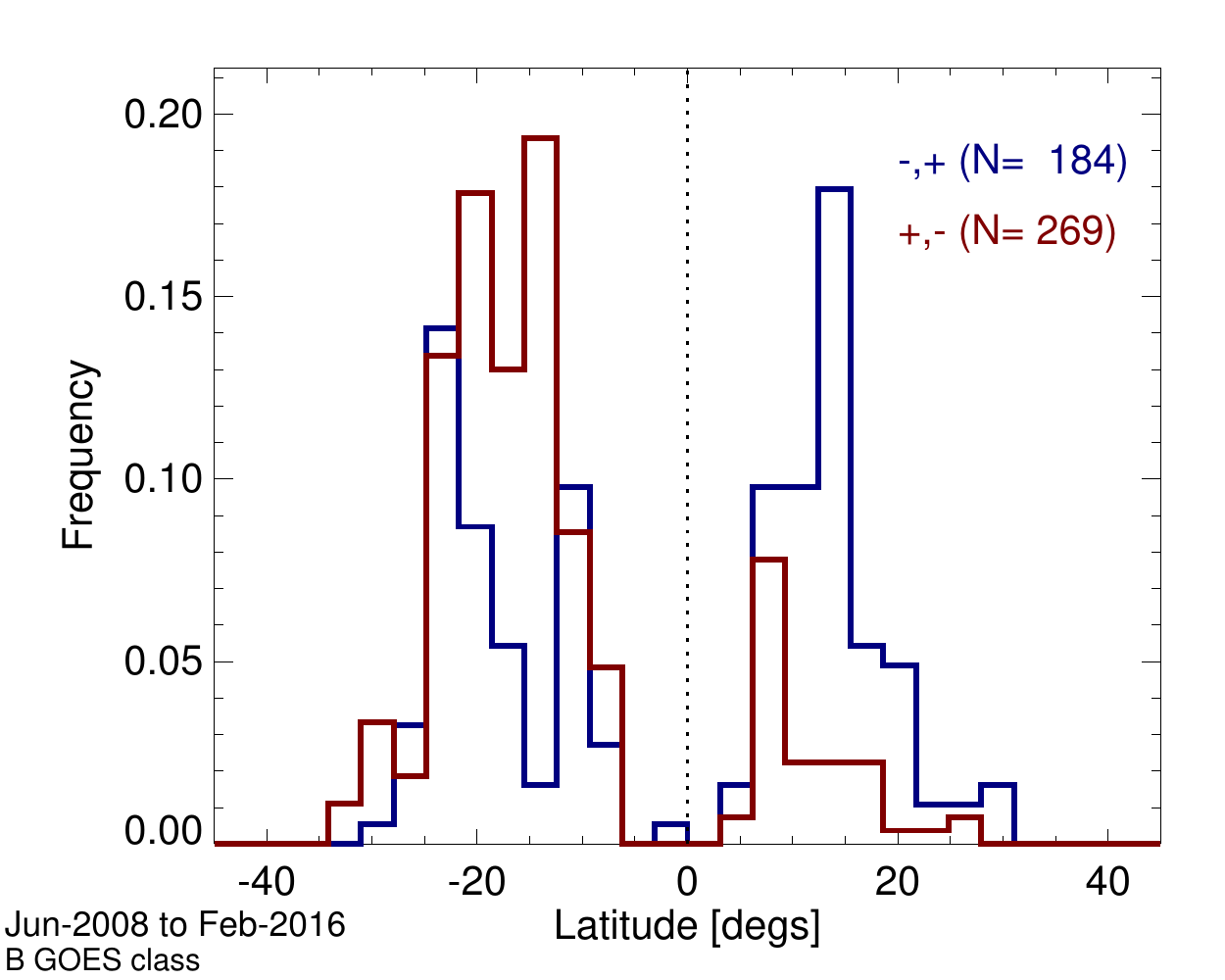}\\
\includegraphics[width=0.4\linewidth]{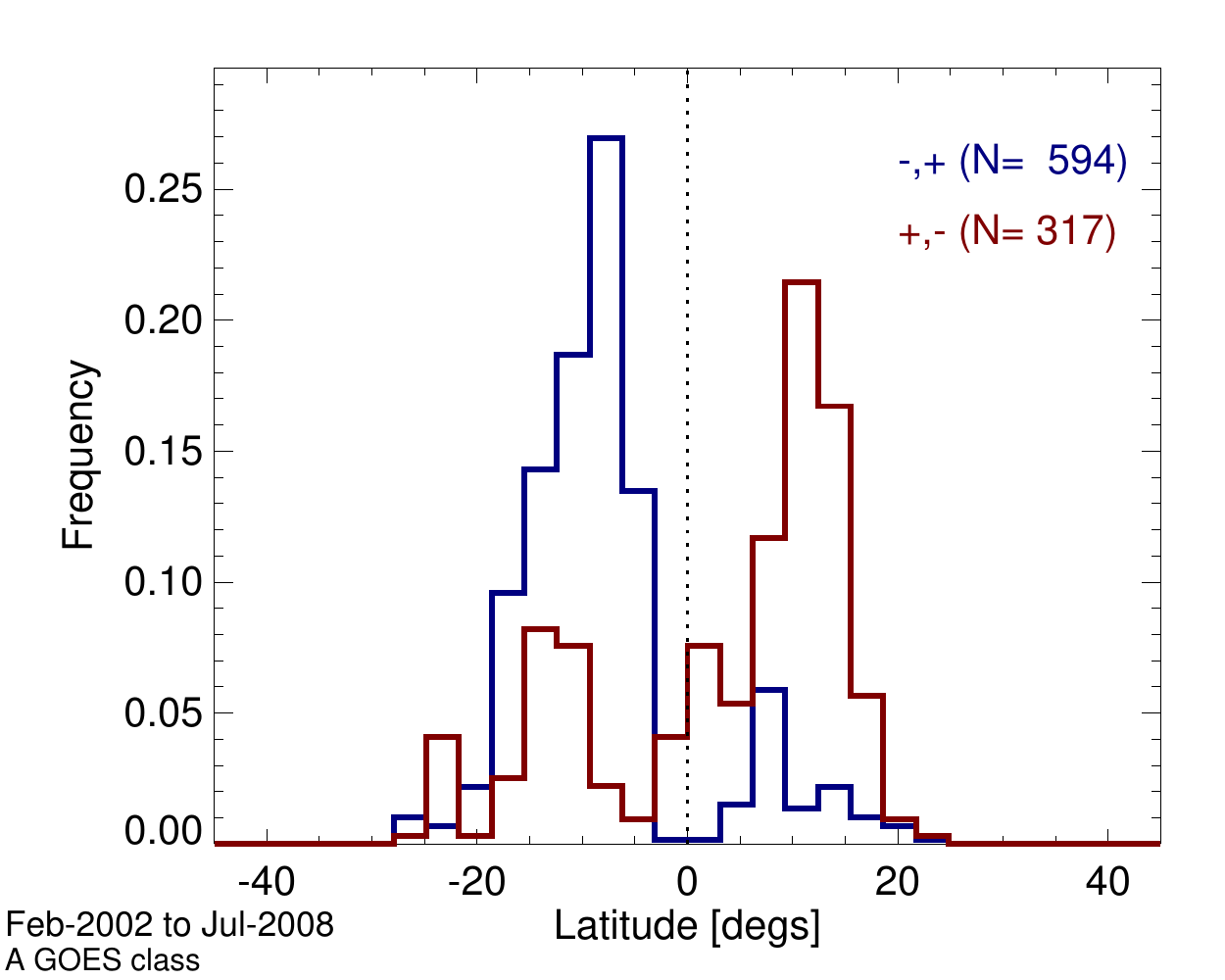}
\includegraphics[width=0.4\linewidth]{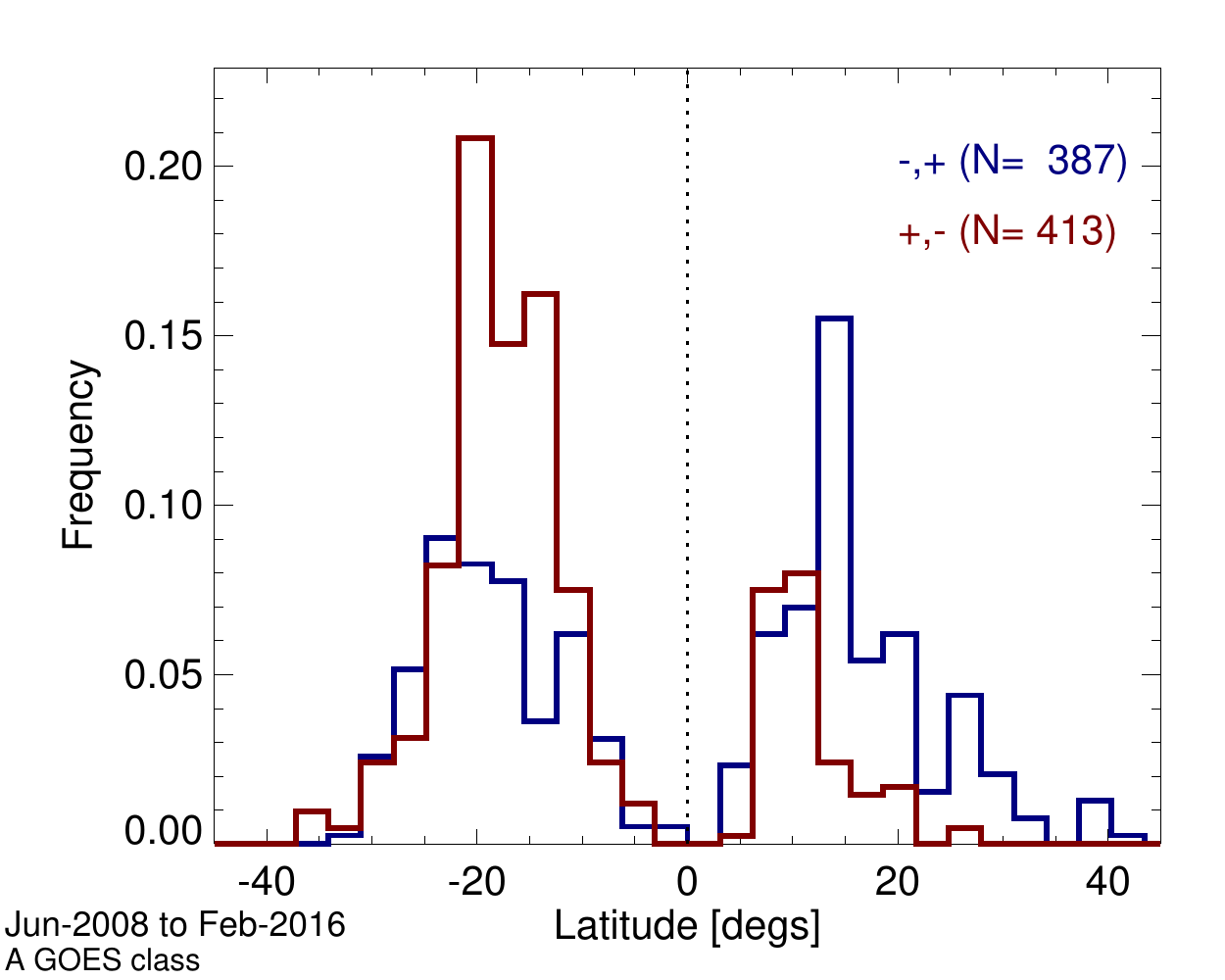}\\
\caption{Latitude histograms of the RHESSI flares with longitudes $|\lambda_{F}|\leqslant25^{\circ}$ that occurred when the HSB was at central meridian. In the left column are data for Cycle 23 and the right for Cycle 24, separated by GOES class (top row X/M-class then C, B, and A-class on the bottom row). The colours indicate the polarity change at the sector boundary, with blue as (-,+) and red as (+,-).}
\label{fig:1d_pergoes}
\end{figure}

Following the methods described by \cite{2011ApJ...733...49S}, we determine the time the HSB was at central meridian at the Sun using the HMF sector boundary crossing detection at the Earth. These times are available online\footnote{\url{http://www.leif.org/research/sblist.txt}} and provide the date of a change from positive to negative (+,-) or negative to positive (-,+) magnetic polarity. With this Earth boundary crossing approach a simple ballistic approximation for the solar wind allows us to determine  when the HSB was at central meridian at the Sun. To take into account the variability of the solar wind speed we assume 4.5 to 6.5 days for propagation. The observed sense of magnetic polarity change determines the hemisphere of the HSB, as shown in Fig.~\ref{fig:hsb_cartoon}. 

We use a modified version of the flare list found by the Reuven Ramaty High Energy Solar Spectroscopic Imager (RHESSI; \citet{2002SoPh..210....3L}), filtering out non-solar or dubious events. The full details of this approach, and resulting checks of the list, are given in Appendix \ref{nsec:rhessifl}. Our filtered list contains 73,711 events over 12 February 2002 (when RHESSI began observations) up to 23 February 2016. These 14 years of flare positions cover the maximum and the declining phase of Cycle 23, through solar minimum and then the rising, maximum and declining phases of Cycle 24. We then consider all the flares that occurred between 4.5 to 6.5 days before each Earth boundary crossing, which reduces our flare list down to 9,189 events. With the restricted list, we then separate the flares that occurred in Cycles 23 and 24 (before and after 1 July 2008) as well as those that occurred during a (+,-) or (-,+) crossing. From these four separate lists we produce 2D histograms of the flare latitude and longitudes, shown in Fig.~\ref{fig:2D_all_flares}.

For Cycle 23, we find that the flares that occurred during the times near central meridian do so preferentially in the hemisphere where the HSB is located (top row Fig. \ref{fig:2D_all_flares}), confirming the results of \citet{2011ApJ...733...49S}. For Cycle 24 (bottom row Fig. \ref{fig:2D_all_flares}) we can immediately see that the association is not as strong but is still present and that the hemispheres for each HSB polarity type have swapped, as expected. For the (+,-) HSB (bottom right Fig.~\ref{fig:2D_all_flares}), there is the expected concentration of flares about the central meridian and they are preferentially in the southern hemisphere, the opposite from Cycle 23. However for the (-,+) HSB (bottom left Fig.~\ref{fig:2D_all_flares}) there are only marginally more events in the northern hemisphere than the southern. Restricting ourselves to only flares with longitudes of $|\lambda_{F}|\leqslant25^{\circ}$, the resulting histograms of flare latitude (shown in Fig. \ref{fig:1D_all_flares}) clearly show more flares in the hemisphere where the HSB is located. The exception is the (-,+) HSB for Cycle 24, which does peak in the expected hemisphere (Northern, blue, right panel in Fig. \ref{fig:1D_all_flares}) but we also find a substantial number in the other hemisphere. 

For each RHESSI flare we can additionally determine the background-subtracted GOES flux (details in Appendix~\ref{nsec:rhessifl}), so we can check whether the HSB association is dependent on flare magnitude. In Fig. \ref{fig:1d_pergoes} we have reproduced the histograms shown in Fig. \ref{fig:1D_all_flares} but separated the events by magnitude going from large flares (X/M-class) through to microflares (A/B-class). In all cases, there is a concentration of events in the hemisphere where the HSB is located. Even with the poorer statistics of the less frequent large flares, we still get the expected pattern.


\section{HSB-Earth and Active Longitude Association}\label{nsec:hsbvsal}

\begin{figure*}
\centering
\includegraphics[width=0.42\linewidth]{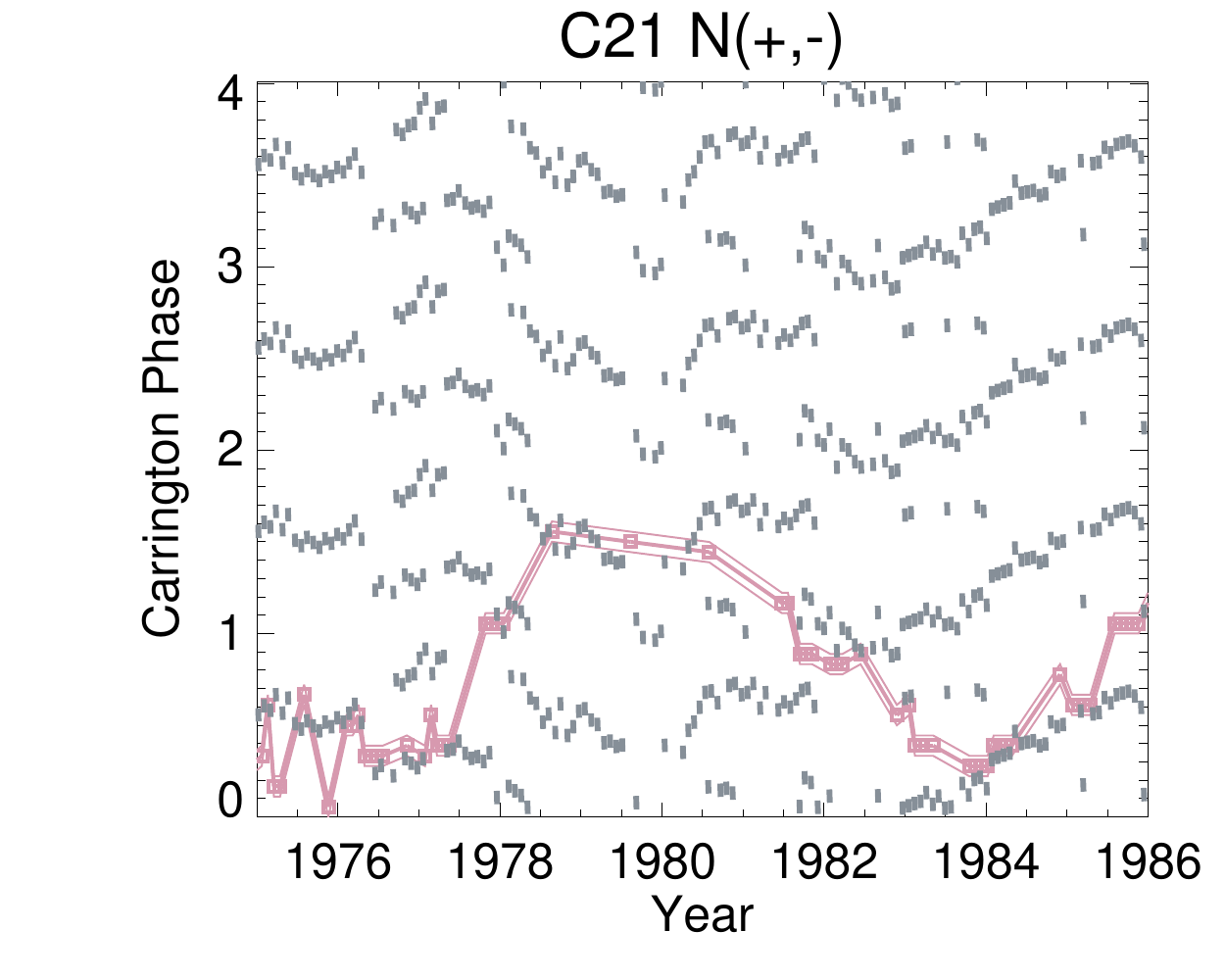}
\includegraphics[width=0.42\linewidth]{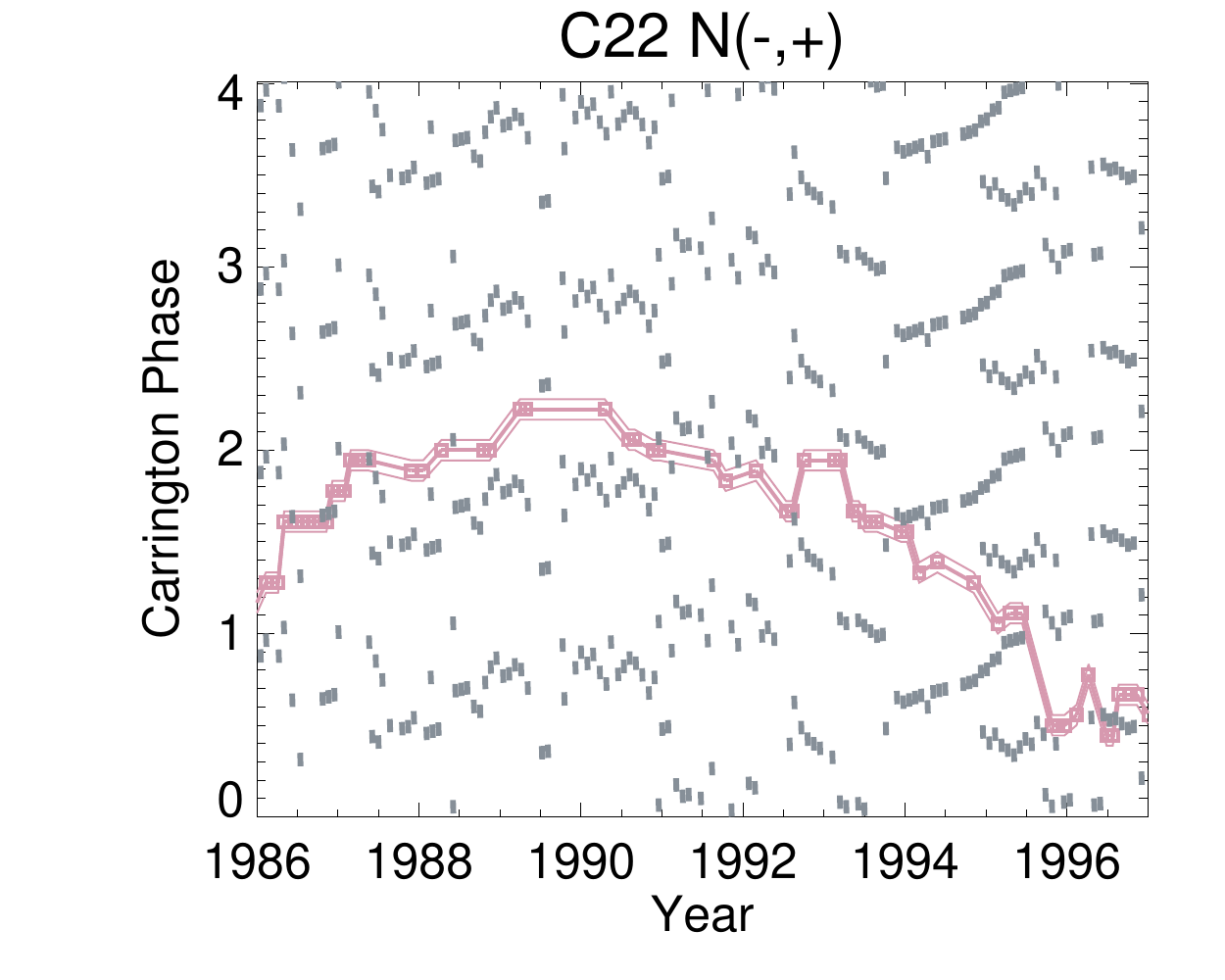}
\includegraphics[width=0.42\linewidth]{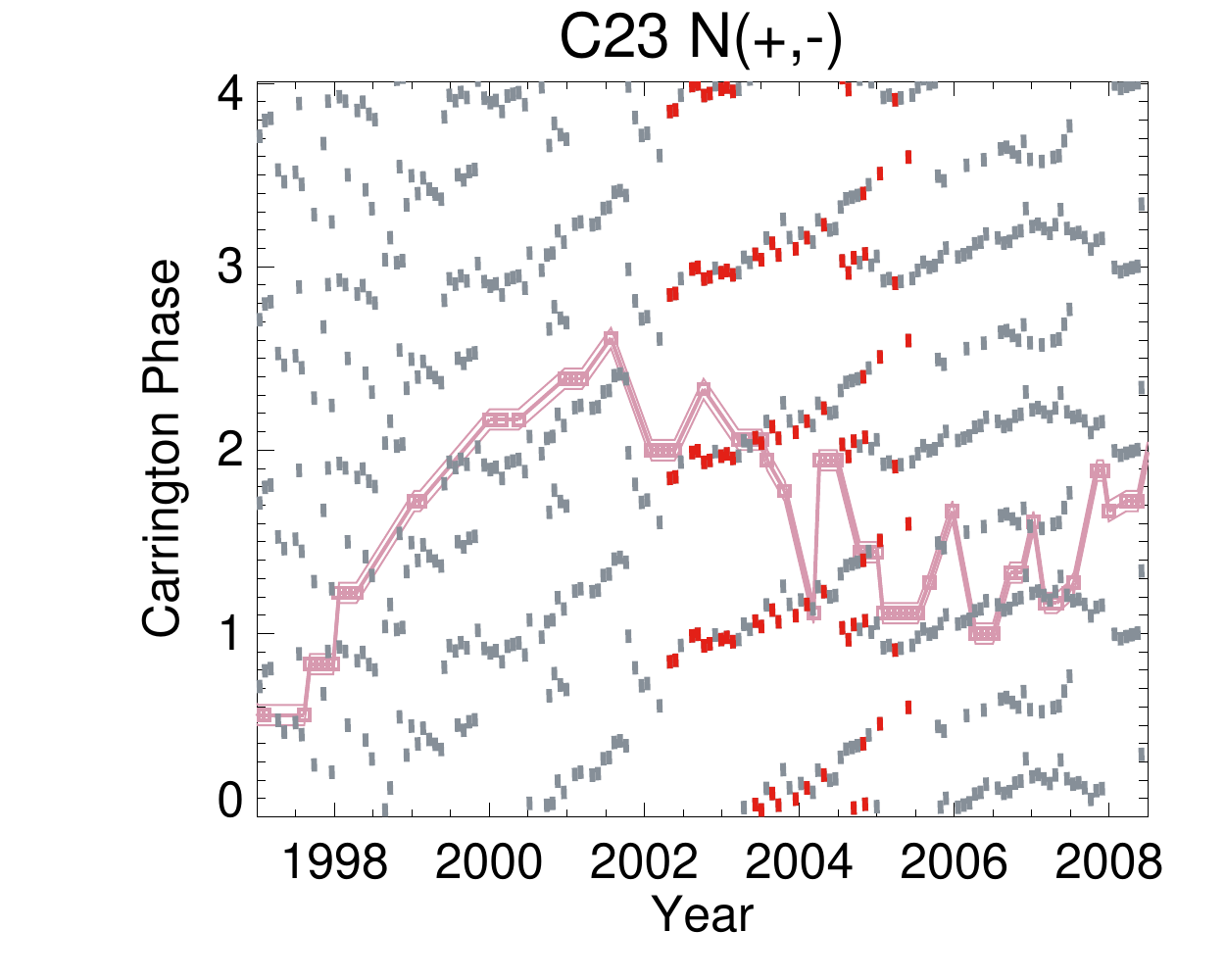}
\includegraphics[width=0.42\linewidth]{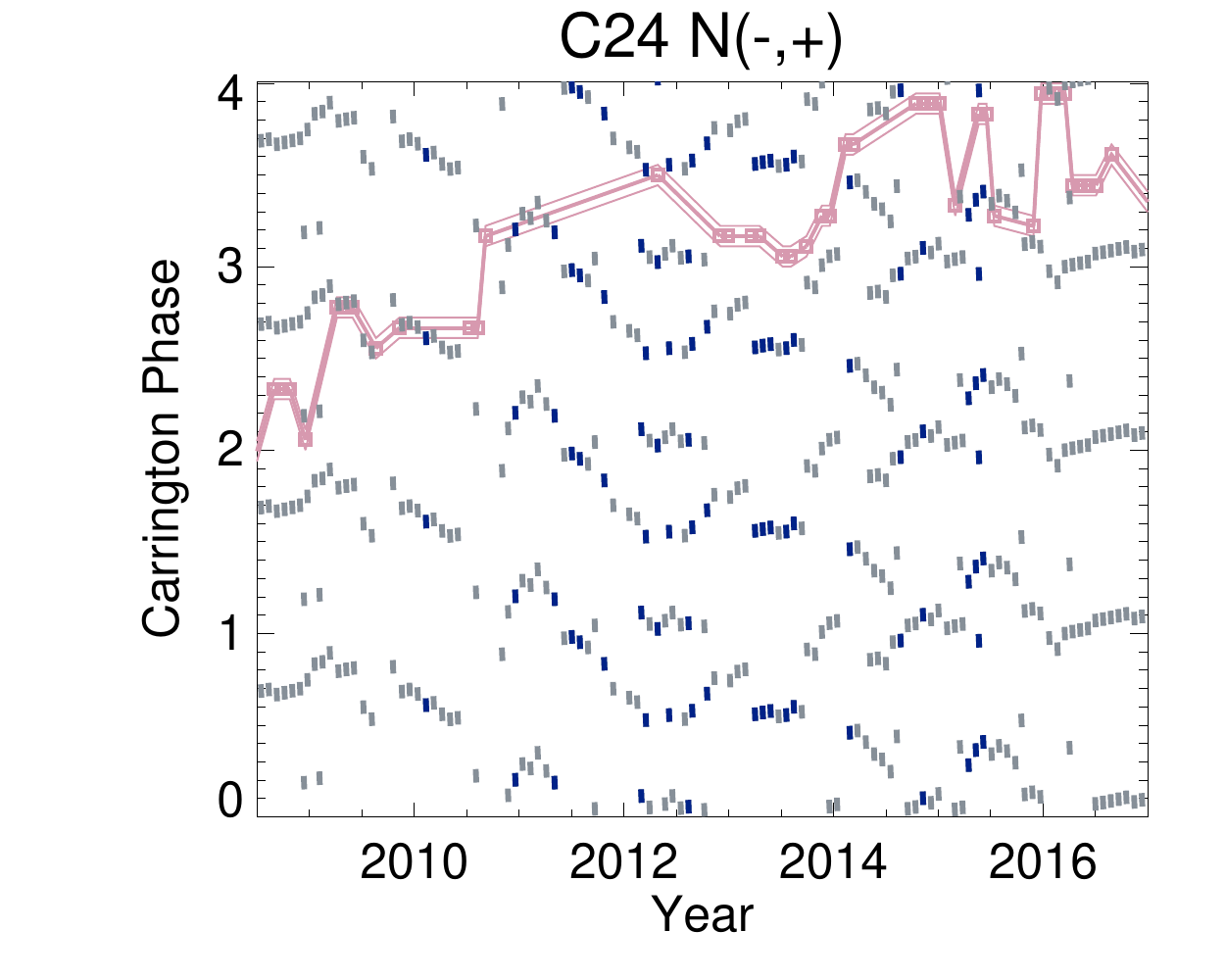}
\caption{Carrington phase against year for the HSB at central meridian over Cycles 21 to 24 (top left to bottom right) in the northern hemisphere. The HSBs which had at least one RHESSI flare occuring nearby (within $\pm 25^\circ$ longitude) are coloured instead of grey, based upon the polarity change, either (+,-) red and (-,+) blue. Also shown are the observed active longitude positions (pink squares) adapted from \citet{2016ApJ...818..127G}.}
\label{fig:cr_vs_cl_2324_n}
\end{figure*}

\begin{figure*}
\centering
\includegraphics[width=0.42\linewidth]{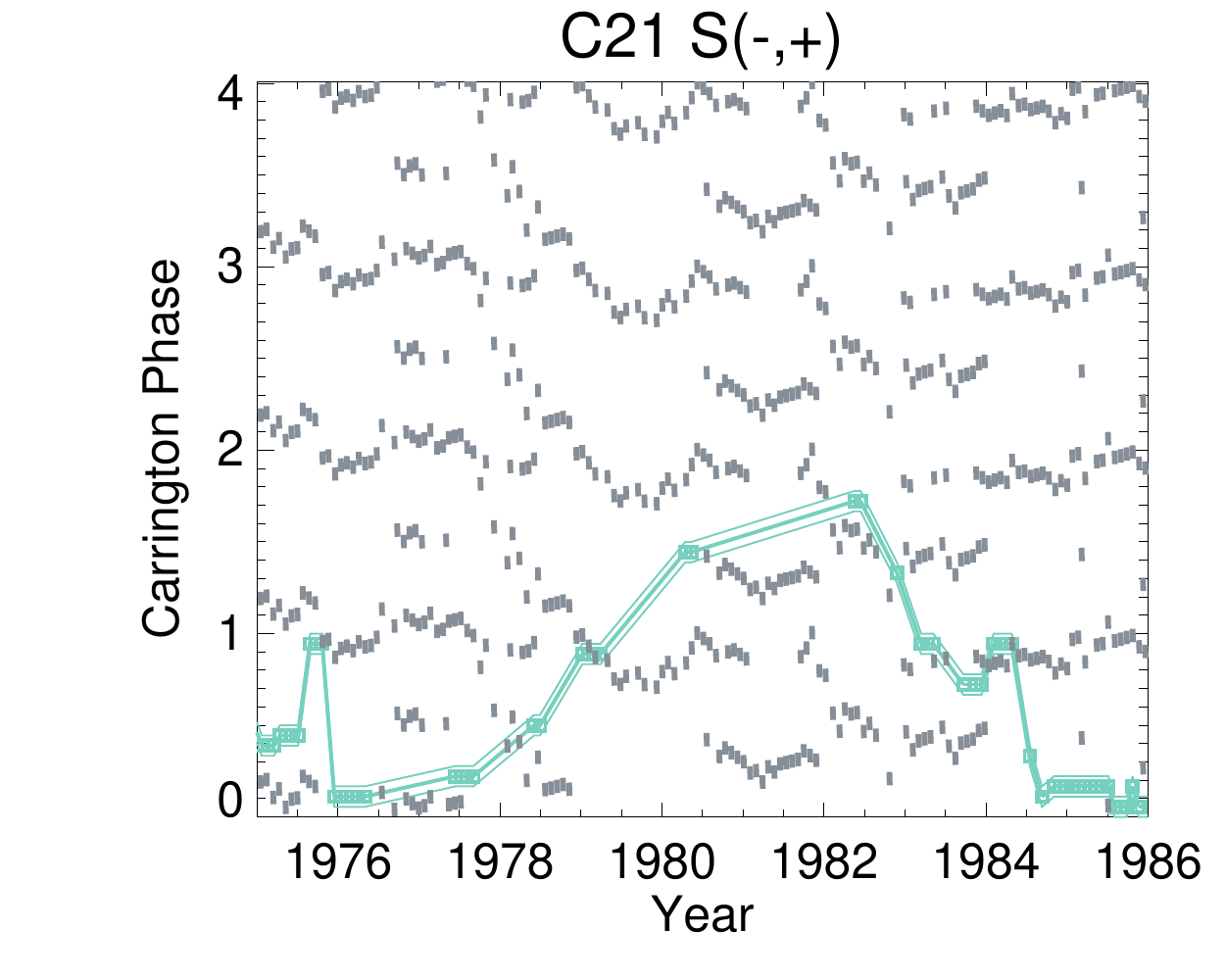}
\includegraphics[width=0.42\linewidth]{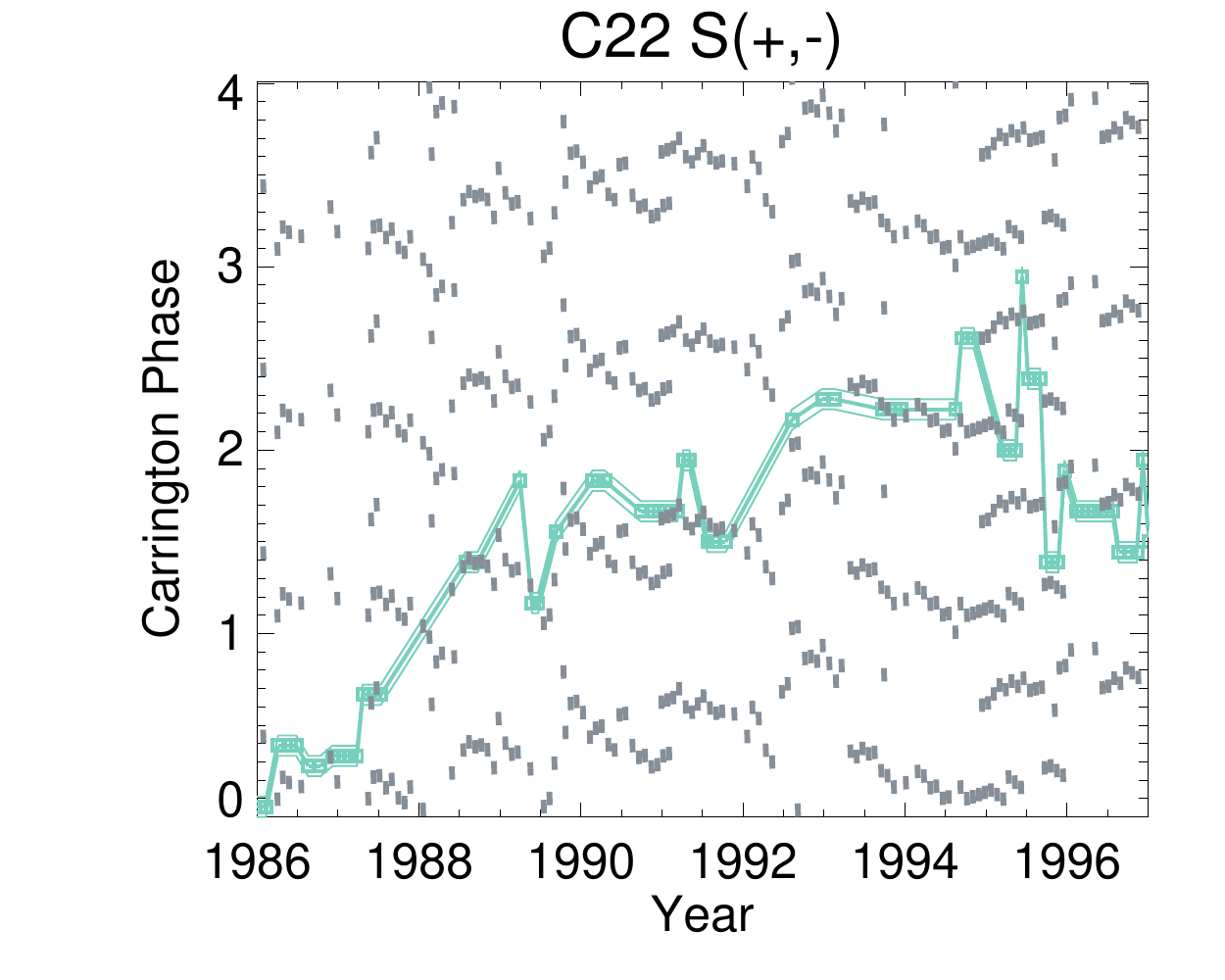}
\includegraphics[width=0.42\linewidth]{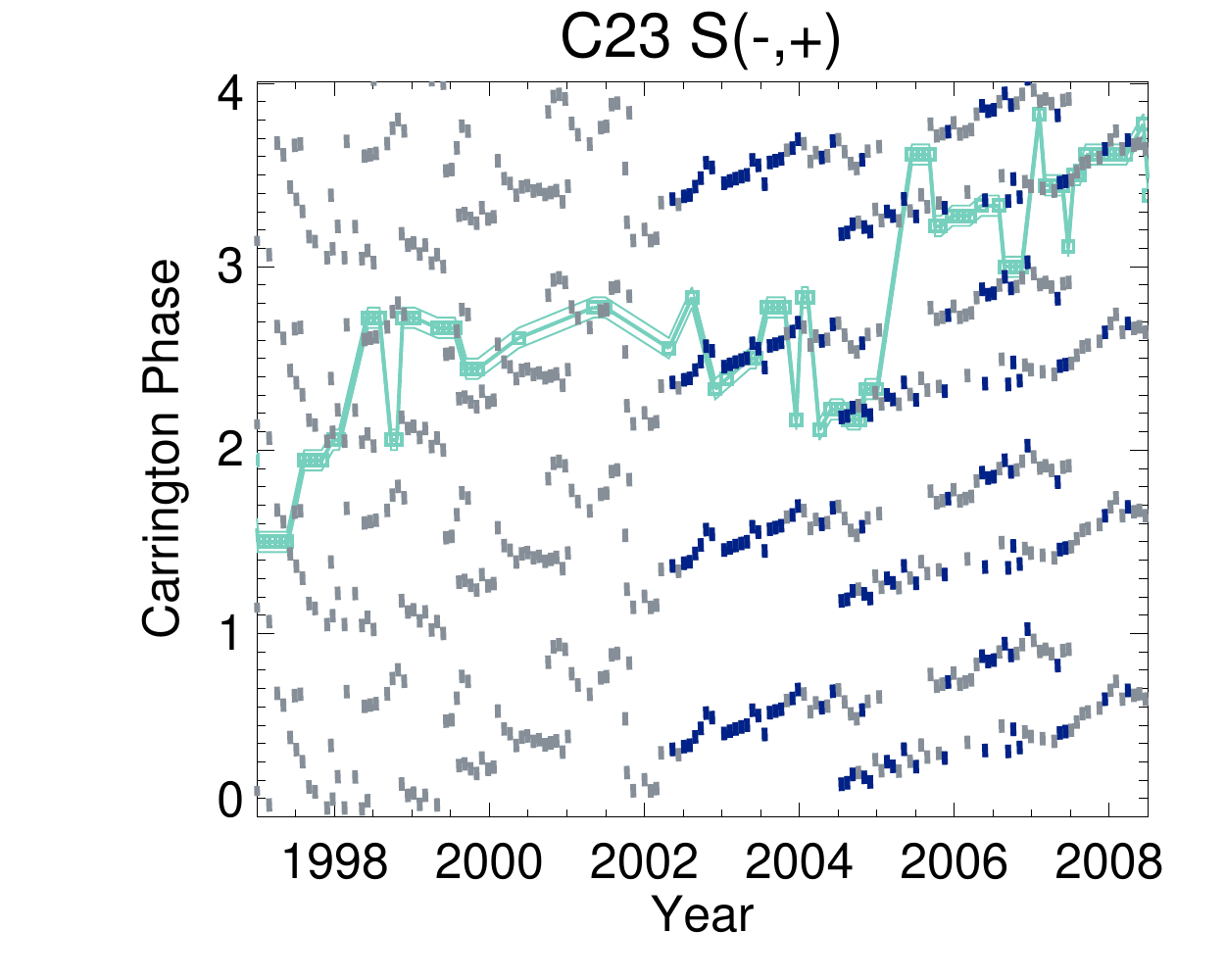}
\includegraphics[width=0.42\linewidth]{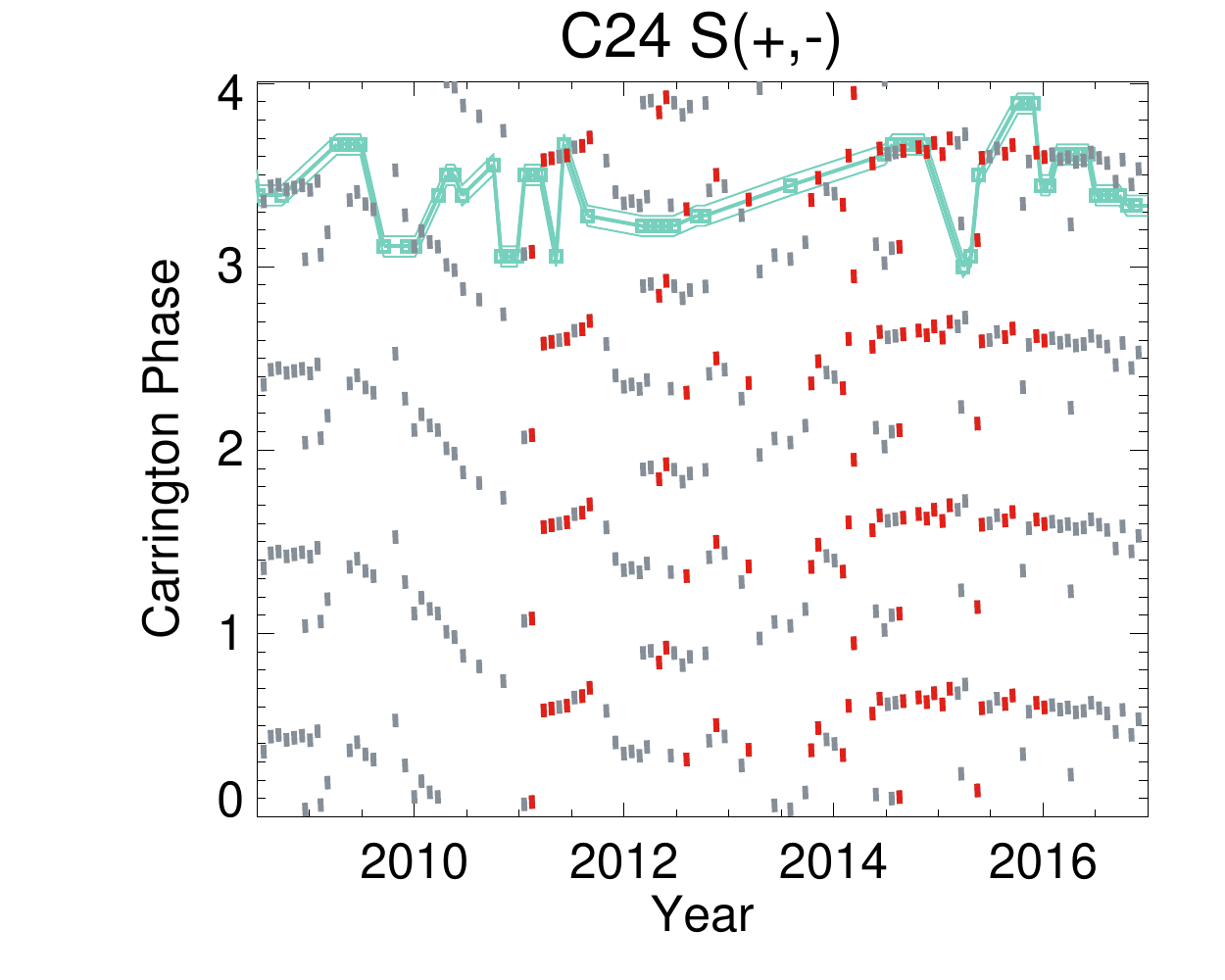}
\caption{Carrington phase against year for the HSB at central meridian over Cycles 21 to 24 (top left to bottom right) in the southern hemisphere. The HSBs which had at least one RHESSI flare occuring nearby (within $\pm 25^\circ$ longitude) are coloured instead of grey, based upon the polarity change, either (+,-) red and (-,+) blue. Also shown are the observed active longitude positions (turquoise squares) adapted from \citet{2016ApJ...818..127G}.}
\label{fig:cr_vs_cl_2324_s}
\end{figure*}

\begin{figure}
\centering
\includegraphics[width=0.8\linewidth]{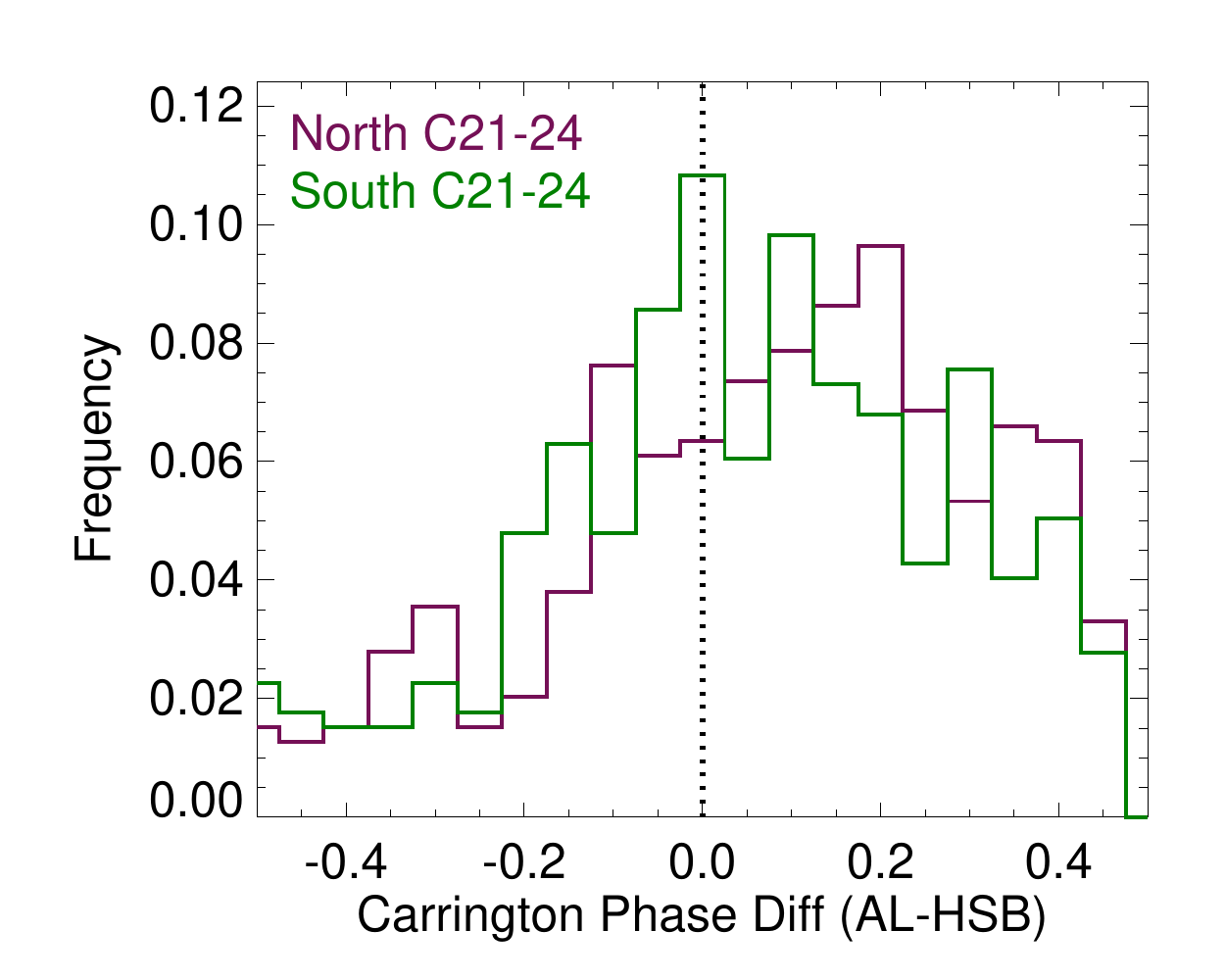}
\caption{Histograms of the Carrington phase difference between the active longitudes and HSBs for the northern (purple) and southern (green) hemispheres.}
\label{fig:alhab_diff_cr}
\end{figure}

We compare our HSB-Earth positions to the active longitudes found by \citet{2016ApJ...818..127G}. These were determined via sunspot observations in the Debrecen photoheliographic data \citep{2016SoPh..291.3081B,2017MNRAS.465.1259G}, and crucially they are given without a differential rotation being applied (private communication Gyenge 2017). These data have a weighted concentration of sunspot group area by Carrington rotation, with a high-pass filter applied to remove weaker and noisier regions. This can result in some Carrington rotations having no identified active longitudes as they are too faint or weak. The Carrington phases, i.e. Carrington longitude$/360^\circ$, of the active longitudes are calculated. The Carrington phases for three additional whole rotations are also plotted so that the path of the dominant active longitude, relative to the Carrington frame over several rotations, can be identified. Repeated regions outside of this path are removed.

The resulting migration paths of the active longitudes over Cycles 21 to 24 are shown in Figs. \ref{fig:cr_vs_cl_2324_n} and \ref{fig:cr_vs_cl_2324_s}, for the northern and southern hemispheres respectively, as points connected by a line with an uncertainty region of $20^\circ$ to account for the longitudinal bins used. Over a solar Cycle, these generally show a ``parabolic migration path'' \citep{2016ApJ...818..127G}, with the phase initially increasing with time (due to rotation faster than the Carrington rate) before levelling out and then decreasing (due to rotating with and then slower than the Carrington rate). On top of this we can also see the ``flip-flop'' behaviour (for instance, from about 2004 onwards in Fig. \ref{fig:cr_vs_cl_2324_n}) where the active longitude phase is sharply jumping between two locations. These would be the locations of the two active longitude bands, but appear as a single sharp change as only the most active band in each rotation is shown.

Our HSB-Earth method gives the times the HSB was at central meridian (over a two-day range) in each hemisphere. We determine the Carrington phase of central meridian at these times to compare to the active longitudes. The resulting Carrington phases of the HSB-Earth are also plotted as a function of year in Figs. \ref{fig:cr_vs_cl_2324_n} and \ref{fig:cr_vs_cl_2324_s}, for the northern and southern hemispheres respectively.  There is by no means a systematic match between the active longitudes and HSB-Earth but there are some periods of rather exact coincidence. This is not surprising since we expect the HSB to be associated with activity, of which the active longitudes are another representation. This association is not improved if we only consider the HSB that actually had RHESSI flares occur nearby (within $\pm 25^\circ$ longitude), shown as the coloured instead of grey points in Figs. \ref{fig:cr_vs_cl_2324_n} and \ref{fig:cr_vs_cl_2324_s}. In particular, during Cycle 24 in the northern hemisphere there are several times with flare-associated HSB that are nowhere near the dominant active longitude (see about the vicinity of 2013 in Fig. \ref{fig:cr_vs_cl_2324_n}).

The HSB drift relative to the Carrington rate in a different manner to the migration path of the active longitudes. The HSBs generally show a Carrington phase that is increasing with time, showing that they are rotating faster than the Carrington rate. The phases in both hemispheres for Cycle 24 show a far flatter path (so rotating with the Carrington rate) but the active longitudes in the Southern hemisphere also show a flatter migration path (certainly less parabolic) for this Cycle.

The HSB phases show the expected pattern of some periods with only one HSB per hemisphere for each rotation (so two sectors overall), and others with two HSBs per hemisphere (four sectors overall). A clear example of the transition between the two and four sectors times is shown about mid-2004 (Cycle 23) in the southern hemisphere, see Fig. \ref{fig:cr_vs_cl_2324_s}. It is generally during these four-sector times that we see the ``flip-flop'' behaviour of the active longitudes, jumping between the two HSB positions (such as from mid-2004 in Fig. \ref{fig:cr_vs_cl_2324_n}).

This association between the active longitudes and HSB in Figs. \ref{fig:cr_vs_cl_2324_n} and \ref{fig:cr_vs_cl_2324_s} could be deceptive as there are numerous times in which no active longitudes have actually been identified. To quantify this association we could consider the Carrington phase difference between the active longitudes and HSB, looking for concentrated peaks. However this is difficult as only small subset of the active longitudes were observed to occur within a day of the HSB being at central meridian (over the four Cycles this was 13\% of the active longitudes in the southern hemisphere and 21\% in the northern). The resulting noisy histogram of the phase difference showed no clear concentration of events, and particularly no peaks at small phase differences. If we instead assume that the path of the active longitudes linearly progresses between detections, we can interpolate these observed locations onto the times at which the HSB were observed at central meridian. Note that for times in which there are two HSB present per hemisphere, we use the one with the smallest phase difference to the interpolated active longitude. The resulting histogram of the Carrington phase difference between the active longitudes interpolated to the HSB times is shown in Fig. \ref{fig:alhab_diff_cr}. There is an association between the active longitudes and HSB in the southern hemisphere, with the largest peaks at small phase differences. However things are more complicated as there are also peaks at non-zero phase differences. The association in the northern hemisphere is not as strong, and the phase difference peaks at about 0.2, suggesting that there is some association but possibly with a consistent offset. For both hemispheres there are more positive than negative phase differences, which corresponds to the HSB moving faster relative to the Carrington rotation rate than the active longitude. This is consistent with the migration paths as we have interpreted them in Figs. \ref{fig:cr_vs_cl_2324_n} and \ref{fig:cr_vs_cl_2324_s}, with the active longitudes having periods of the phase increasing and decreasing but the HSB generally having the phase increasing. However these differences in phase could be due to one or more of the assumptions used in their determination (constant solar wind speed for the HSB, linear interpolation between the observed active longitude positions, only using the dominant active longitude position) and not indicative of a physically real link between them.


\section{HSB determined by PFSS and Comparisons}\label{nsec:findhsbpfss}
\subsection{Method}\label{subsec_method}

\begin{figure*}
\centering
\includegraphics[width=0.35\linewidth]{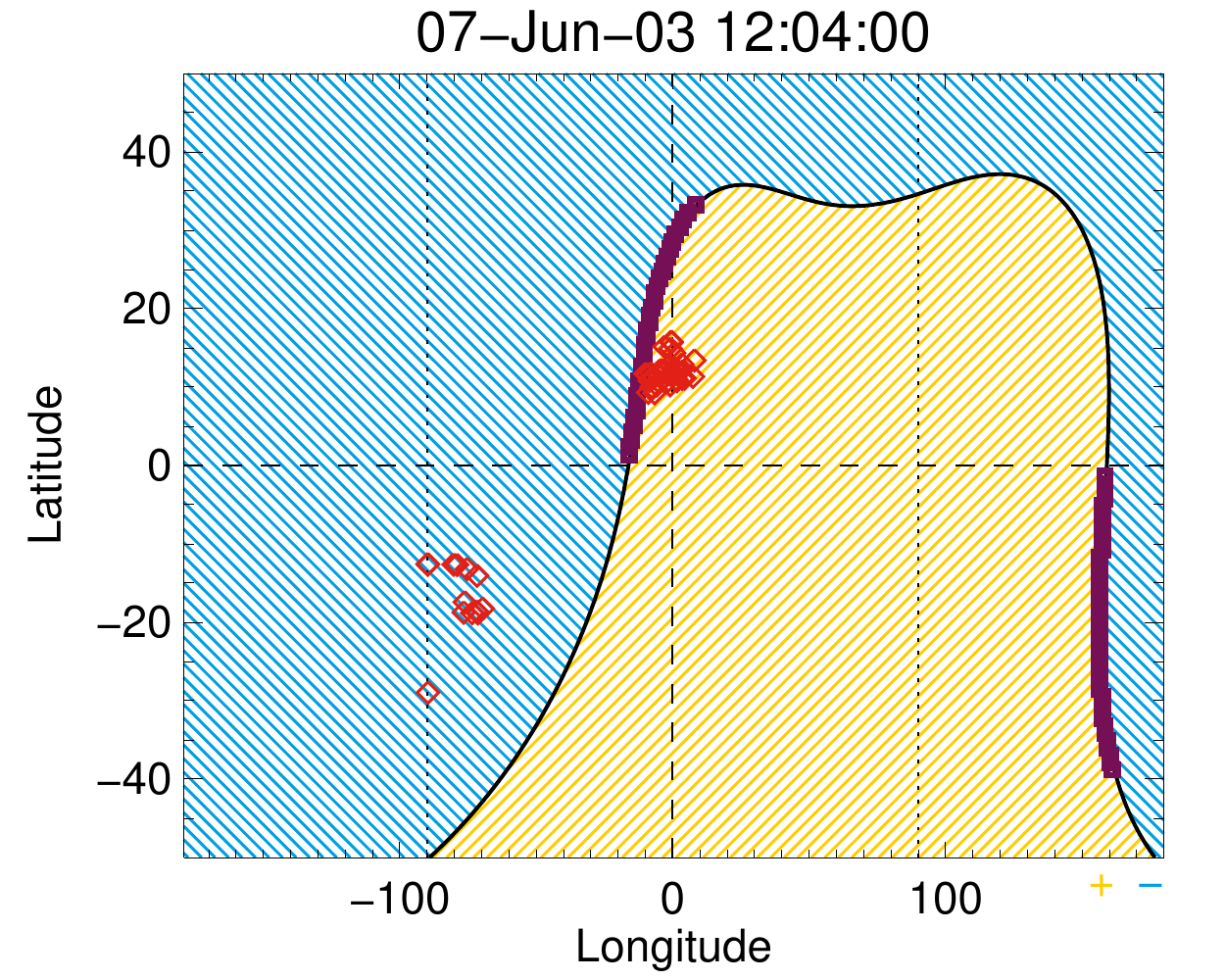} 
\includegraphics[width=0.35\linewidth]{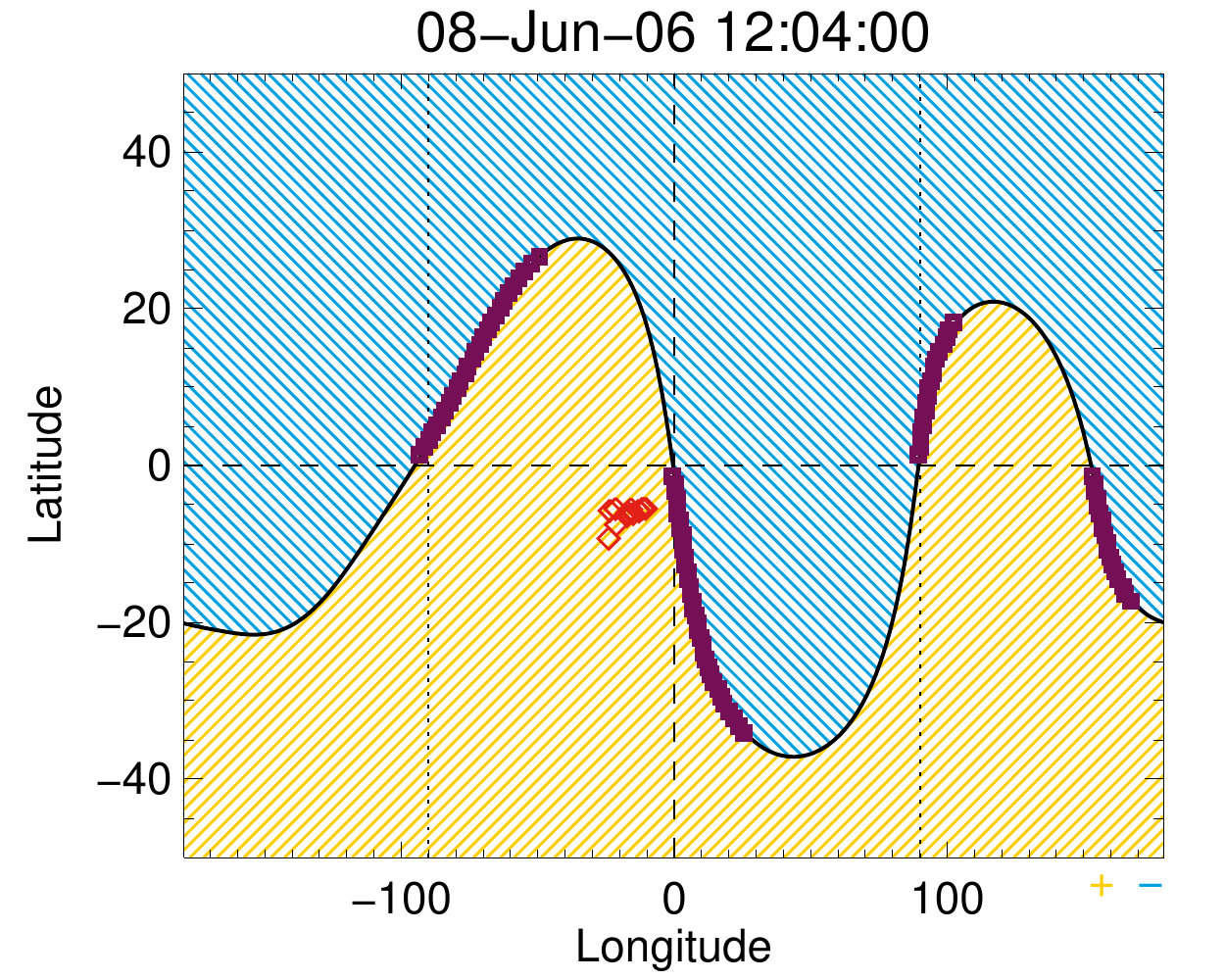}\\
\includegraphics[width=0.35\linewidth]{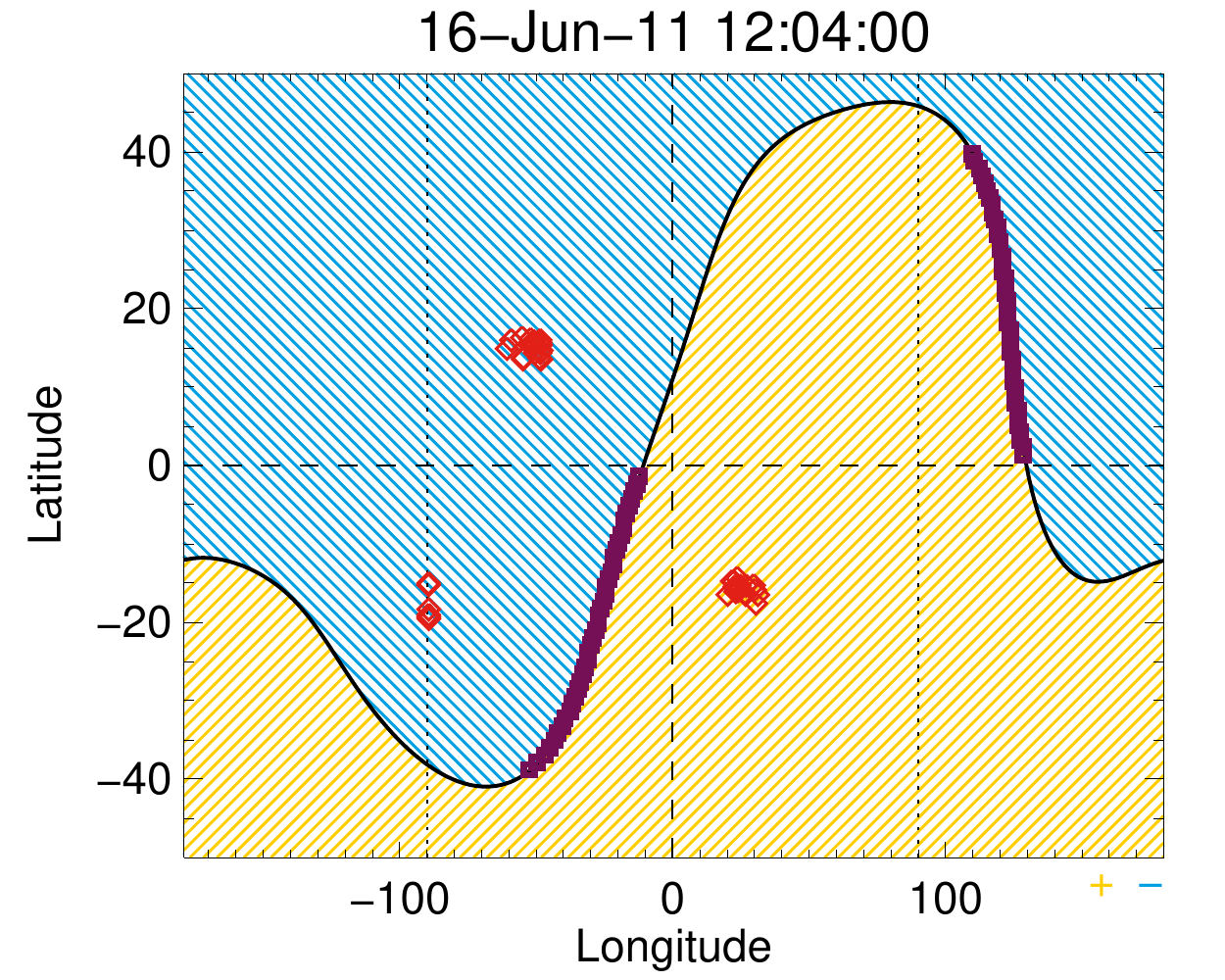}
\includegraphics[width=0.35\linewidth]{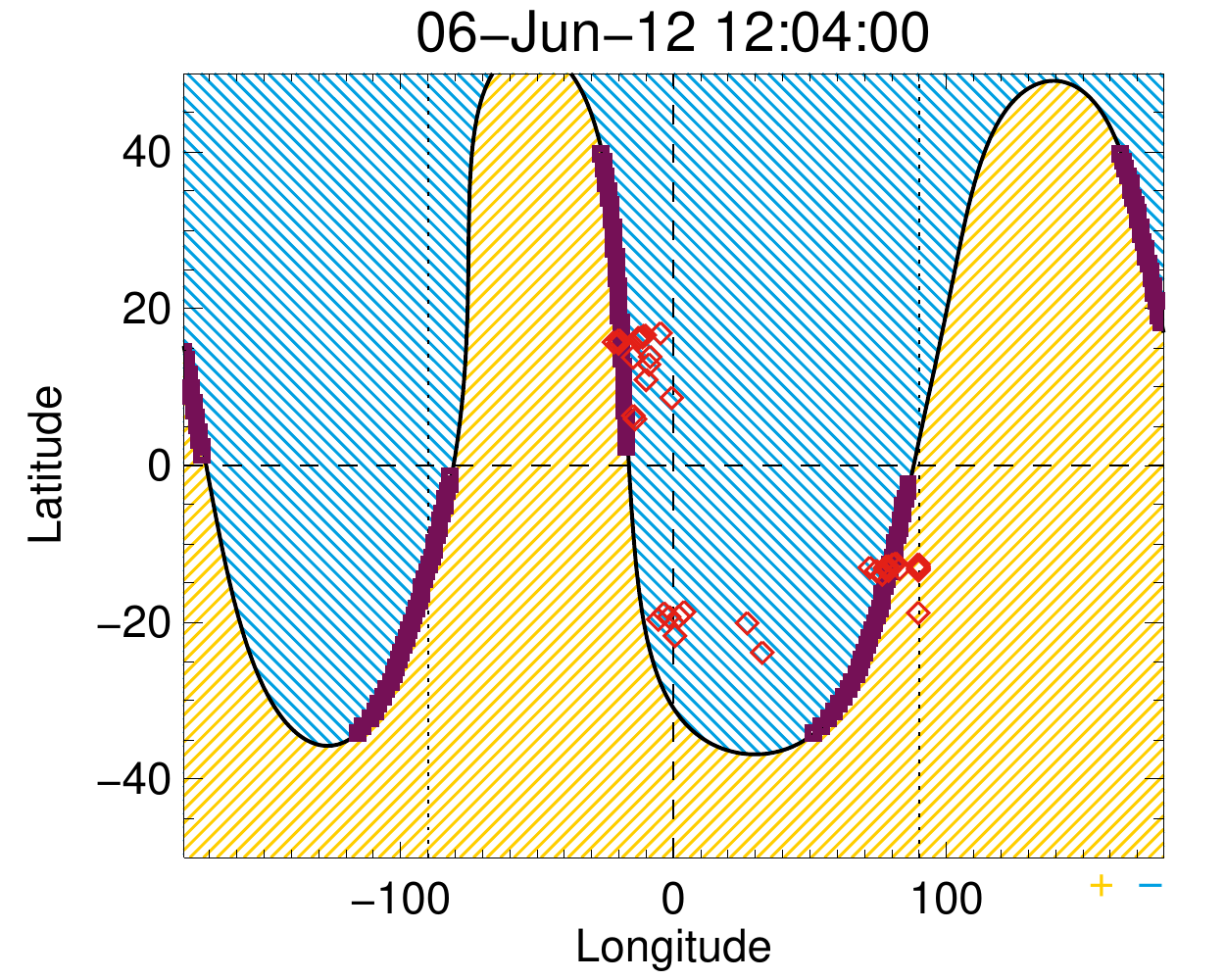}\\
\includegraphics[width=0.35\linewidth]{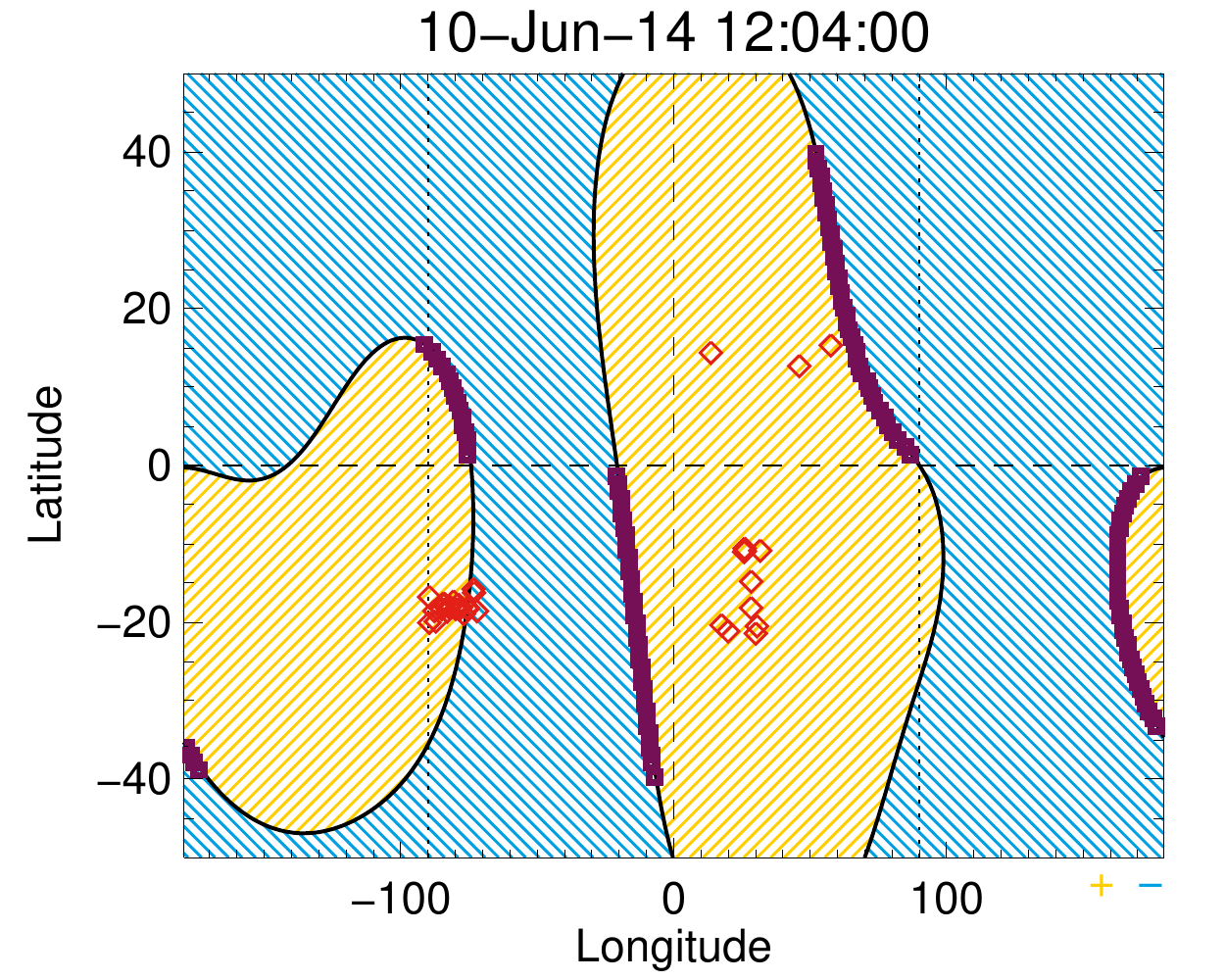}
\includegraphics[width=0.35\linewidth]{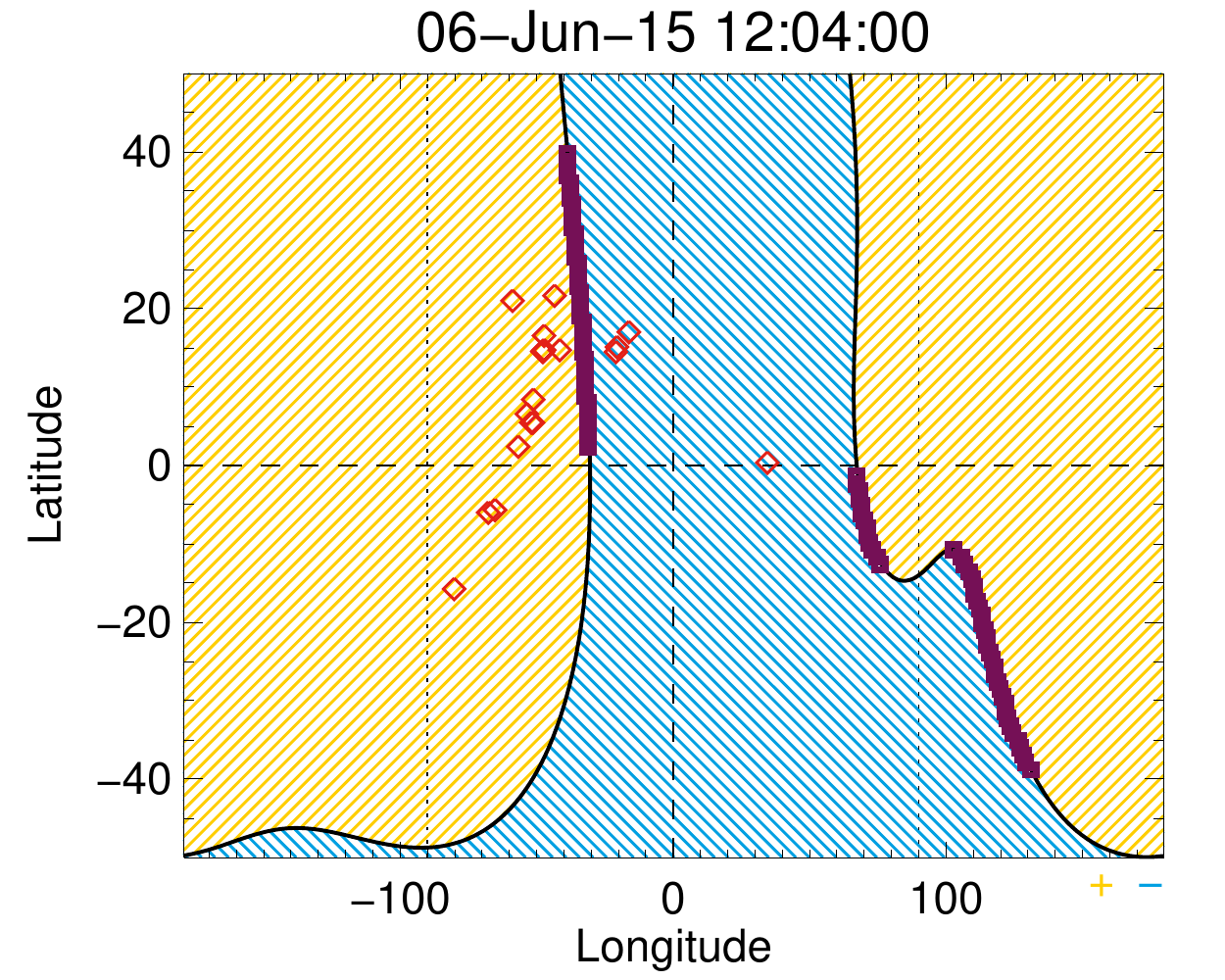}
\caption{Examples of PFSS extrapolations showing the radial magnetic field component at the outermost surface ($2.5R_\sun$) for six different dates, on which the HSB-Earth are predicted to be at central meridian. In the intensity map blue is negative polarity, yellow is positive, with the black line being the sector boundary (neutral line). Over-plotted are the RHESSI flares (red crosses) that occurred within $1$ day. The parts of the sector boundary marked with purple points are the HSB, found out to $\pm40^\circ$ latitude.}
\label{fig:bouter_hsb}
\end{figure*}

Although the HSB-Earth approach in \S\ref{nsec:findhsbesc} is well established and simple, it is dependent on the ballistic assumption about the solar wind speed to map the HSB back to the Sun. It can also only provide the times the HSBs were at central meridian, and without any latitude information. To be able to identify the HSBs at all times, and hence for all flares, we instead develop a new approach using Potential Field Source Surface (PFSS) \citep{1969SoPh....9..131A, 1969SoPh....6..442S} extrapolations of photospheric magnetograms. The model and magnetic data products used in this paper were created by M. DeRosa and K. Schrijver, and available online\footnote{\url{http://www.lmsal.com/~derosa/pfsspack/}} from SOHO/MDI and SDO/HMI (post-2010) magnetograms, generated routinely at 6-hour intervals. Using a daily PFSS at 12:04 UT we take the radial magnetic field component on the outermost surface, for this model at $2.5R_\sun$. In these we can easily identify the location of the neutral line and then for a given latitude determine the longitude at which the magnetic polarity changes in the same way as the leading-to-following sunspots polarity in each hemisphere. Thus we are able to identify the HSB locations in each hemisphere, which we will term HSB PFSS. In this analysis we ignore the time increment between the source surface of the PFSS and the photosphere.

Fig.~\ref{fig:bouter_hsb} shows examples of these PFSS magnetic equators, i.e. the neutral lines of the outer source surface, for days on which the HSB-Earth approach predicted a HSB at central meridian. In the top row of Fig. \ref{fig:bouter_hsb} we show examples from Cycle 23 with two and then four-sector structures (left and right respectively). In both cases shown there is a HSB at central meridian, there are numerous RHESSI flares at these HSBs, and these HSBs are relatively vertical.  In the middle row of Fig. \ref{fig:bouter_hsb} we show examples from Cycle 24, again with the two and then four-sector structures (left and right respectively). As expected, the polarity of the HSBs near central meridian is different in each hemisphere compared to Cycle 23. This time however the HSBs are both slightly before central meridian, which could indicate that the solar wind speed was faster than expected, resulting in a shorter time between HSB at central meridian and polarity reversal detected at the Earth. Again there are concentrations of RHESSI flares near some of the HSBs but there are also several instances of flares far from the HSBs. The bottom row of Fig. \ref{fig:bouter_hsb} shows more complicated examples from Cycle 24. Again there are HSBs near central meridian in the expected hemisphere but not fully rotated around to this location, consistent with a faster solar wind speed. The most extreme case (bottom right Fig. \ref{fig:bouter_hsb}) has the HSB occurring about $30^\circ$ before central meridian, so about two days away from central meridian, but is still in the correct hemisphere; moreover there is a cluster of flares about this HSB. Overall, the examples in Fig.~\ref{fig:bouter_hsb} show the desired relationship, though with some discrepancies, that we quantify further in Section~\ref{nsec:pfssvsfl}.

\subsection{Comparison of the two HSB approaches}\label{nsec:escvspfss}

\begin{figure*}\centering  
\includegraphics[width=0.4\linewidth]{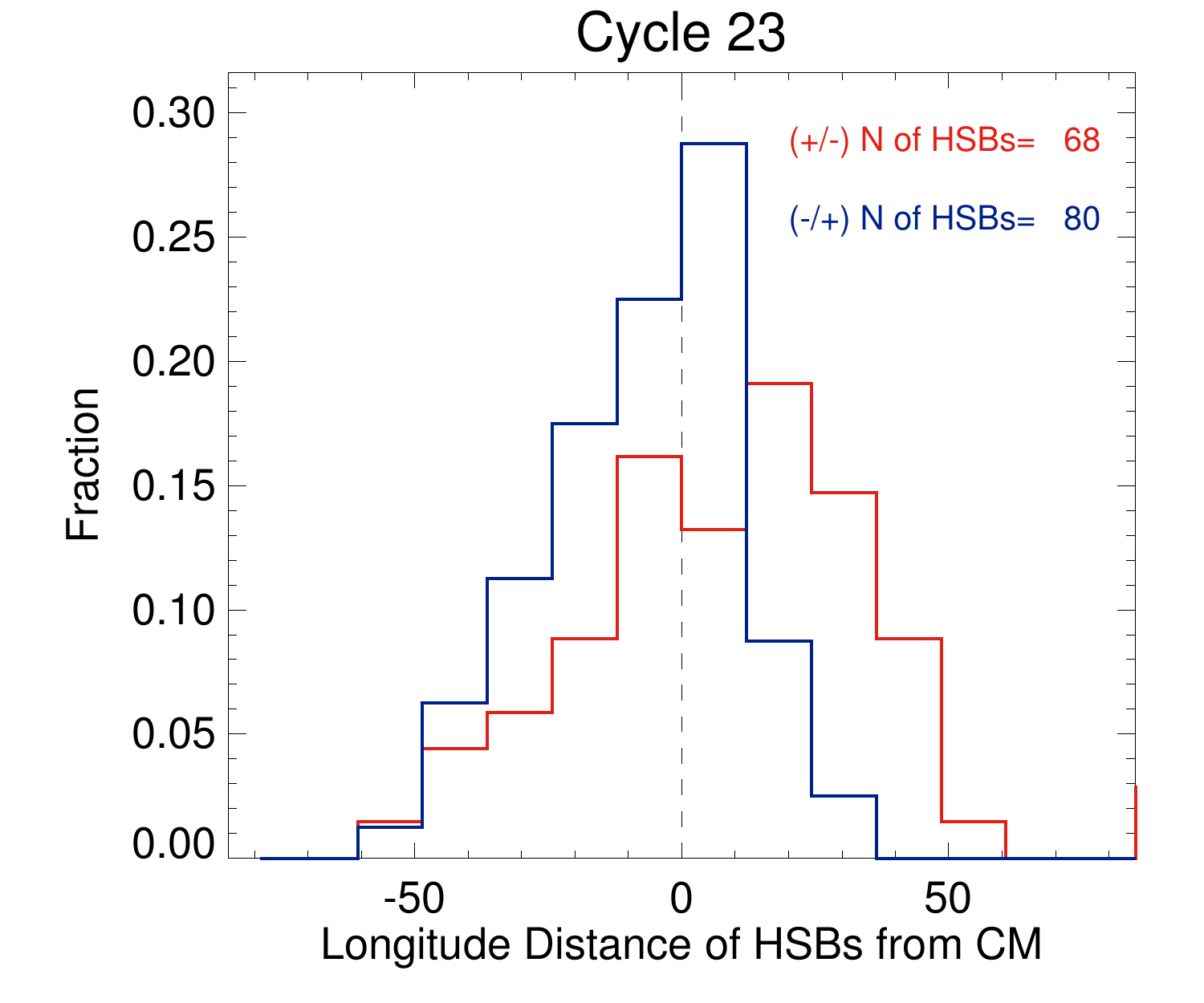}
\includegraphics[width=0.4\linewidth]{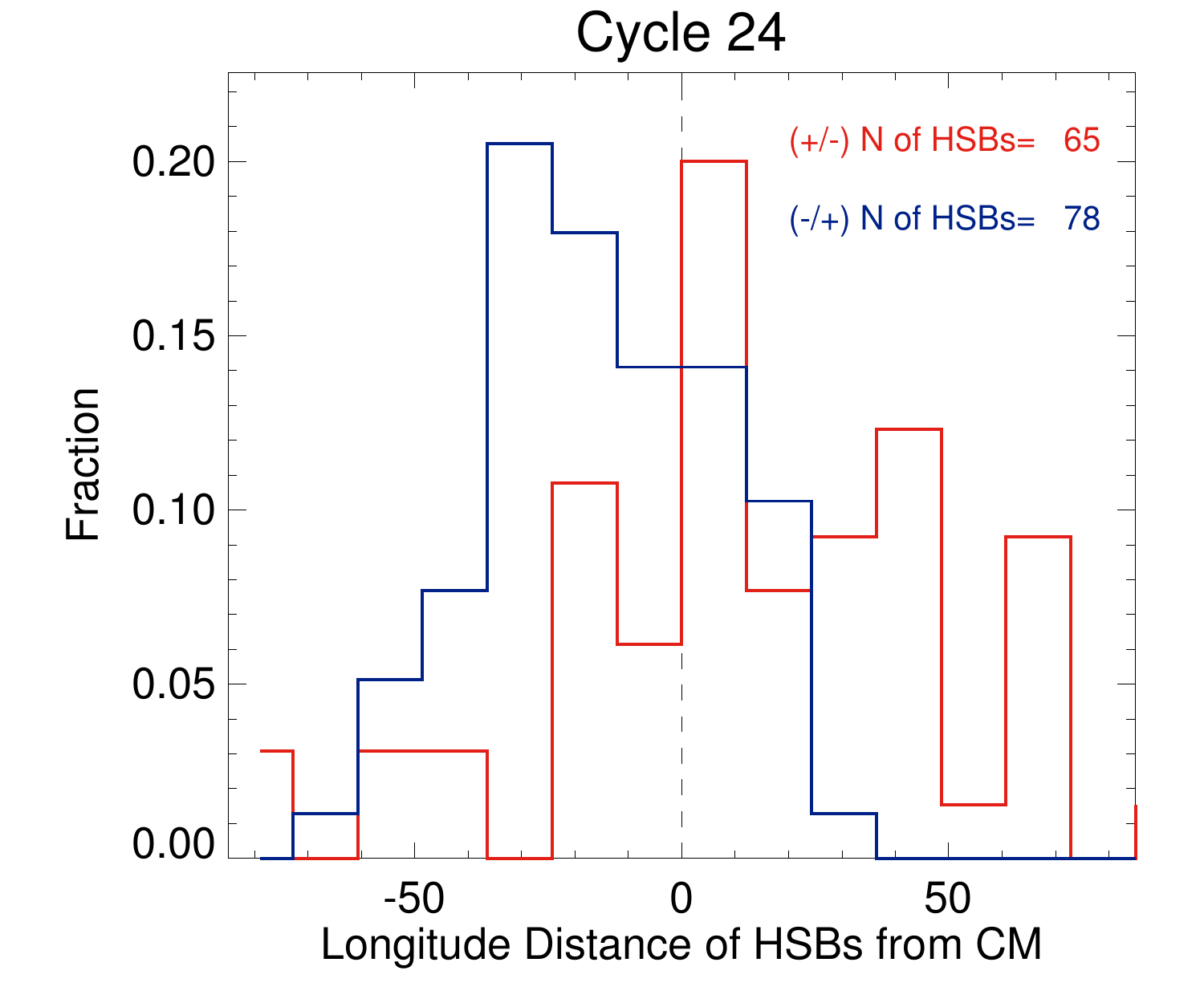}
\caption{Longitudinal distance of the HSB PFSS on the days they were predicted to be at central meridian by the HSB-Earth method, for the days RHESSI observed flares. For Cycle 23,  $89\%$ of HSB PFSS with polarity change (-,+) were no further than $30^o$ away from central meridian, while the percentage for the HSB PFSS (+,-) was $68\%$. For Cycle 24 $68\%$ of HSB PFSS (-,+) were within $30^o$ of central meridian, with $66\%$ for HSB PFSS (+,-).} 
\label{fig:hsb_dist}
\end{figure*}

To check the performance of the two HSB detection methods we determined the HSB PFSS (see \S\ref{subsec_method}) for the days that the HSB-Earth (see \S\ref{nsec:findhsbesc}) predicted they would be at central meridian. We restricted ourselves to days only when RHESSI detected flares, as that is the set of HSBs that are used in the HSB-Earth flare analysis (see \S\ref{nsec:findhsbesc}). We calculated the HSB PFSS at the equator, as the HSB-Earth approach provides no latitude information other than hemisphere. The resulting histograms of the longitude distance of the HSB PFSS from central meridian are shown in Fig. \ref{fig:hsb_dist}, separated by Cycle and HSB polarity change.  For Cycle 23 we find that both peak at or near central meridian, with 89\% within $30^\circ$ for the (-,+) change and 68\% for (+,-). For Cycle 24 the correlation is weaker with 68\% and 66\% respectively for each HSB polarity change. However, for Cycle 24's (-,+) change it is worse, with a clear peak about $30^\circ$ before central meridian. This helps explain the discrepancy we saw for this configuration in Fig. \ref{fig:2D_all_flares} (bottom left) and \ref{fig:1D_all_flares} (right panel) where the HSB-flare association was weaker. It seems that in this case the HSBs would occur about 2 days later (i.e. about 2.5 to 4.5 days before the reversal was detected at the Earth) than we used for the analysis in \S\ref{nsec:findhsbesc}, but mostly for northern events. These results suggest that our approach of determining the HSB from the PFSS works reasonably well and has the potential to sharpen the apparent concentration of flare occurrences, as compared with the Earth-boundary approach. \citet{2017SoPh..292..174G} have used an improvement on the simple HSB-Earth approach, using the measured solar wind speed. However with this refined approach they do not find a substantial difference over the Cycles we study in this paper.

\subsection{HSB PFSS and flare occurrence}\label{nsec:pfssvsfl}

\begin{figure*}\centering  
\includegraphics[width=0.4\textwidth]{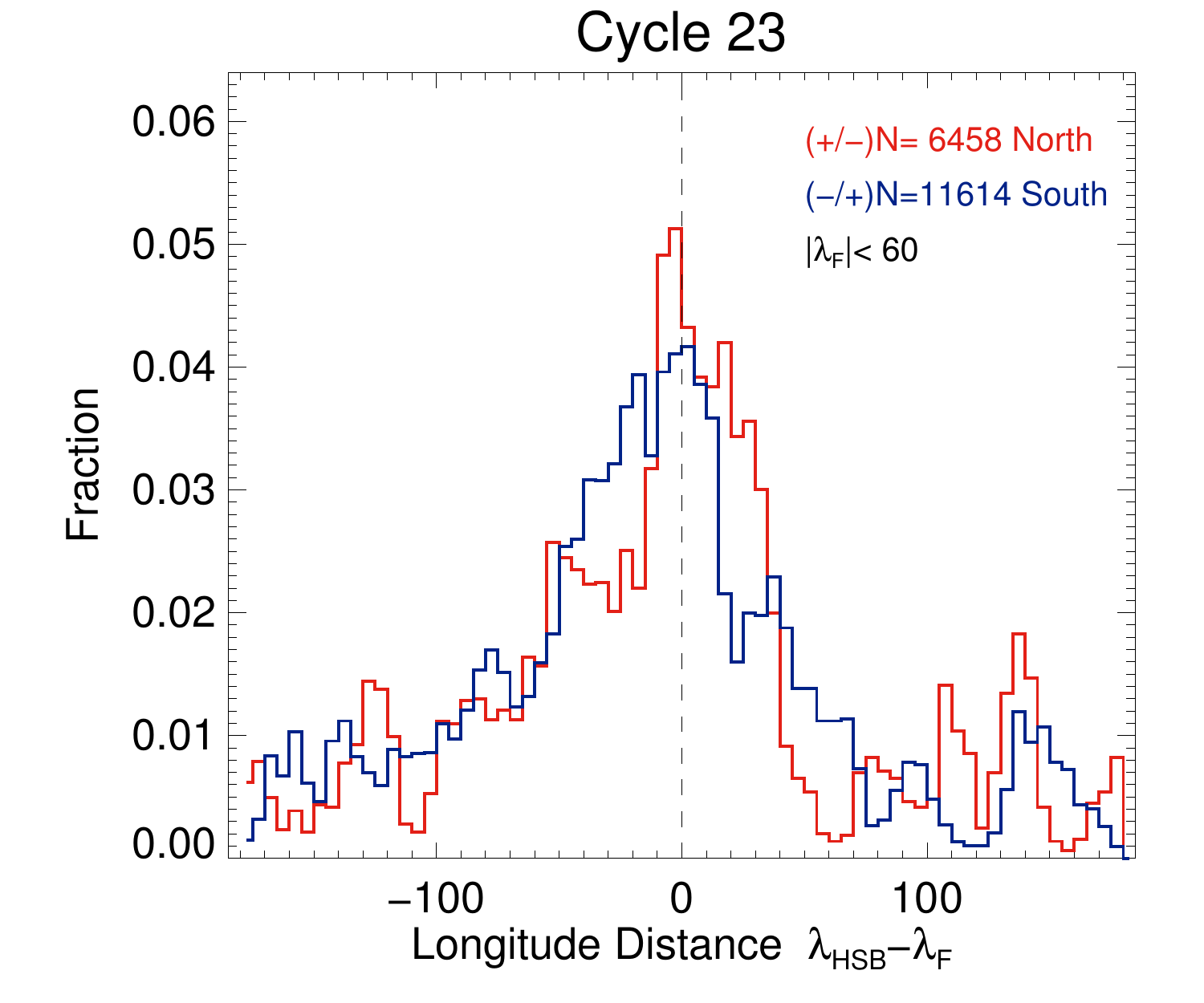}
\includegraphics[width=0.4\textwidth]{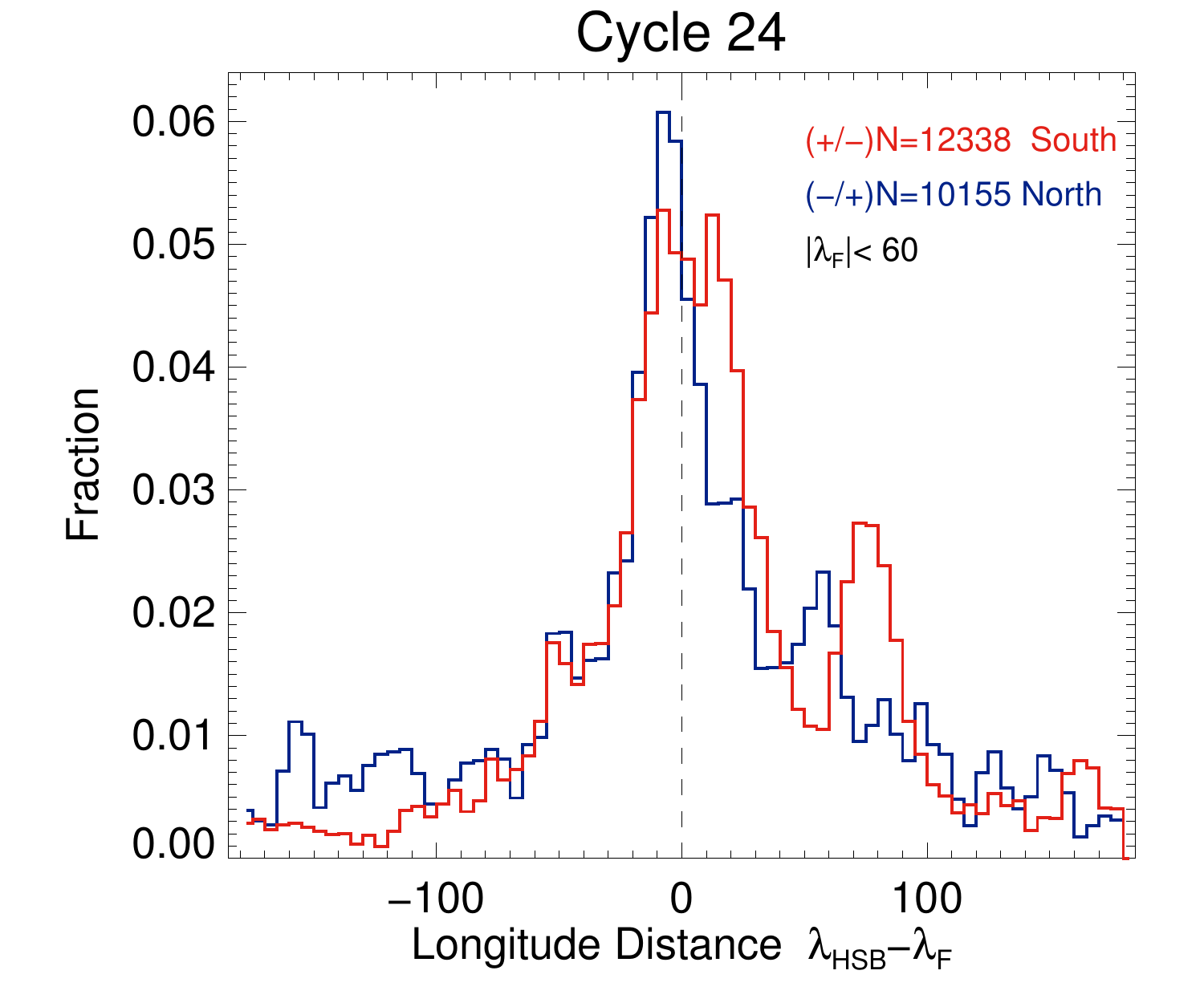}\\
\includegraphics[width=0.4\textwidth]{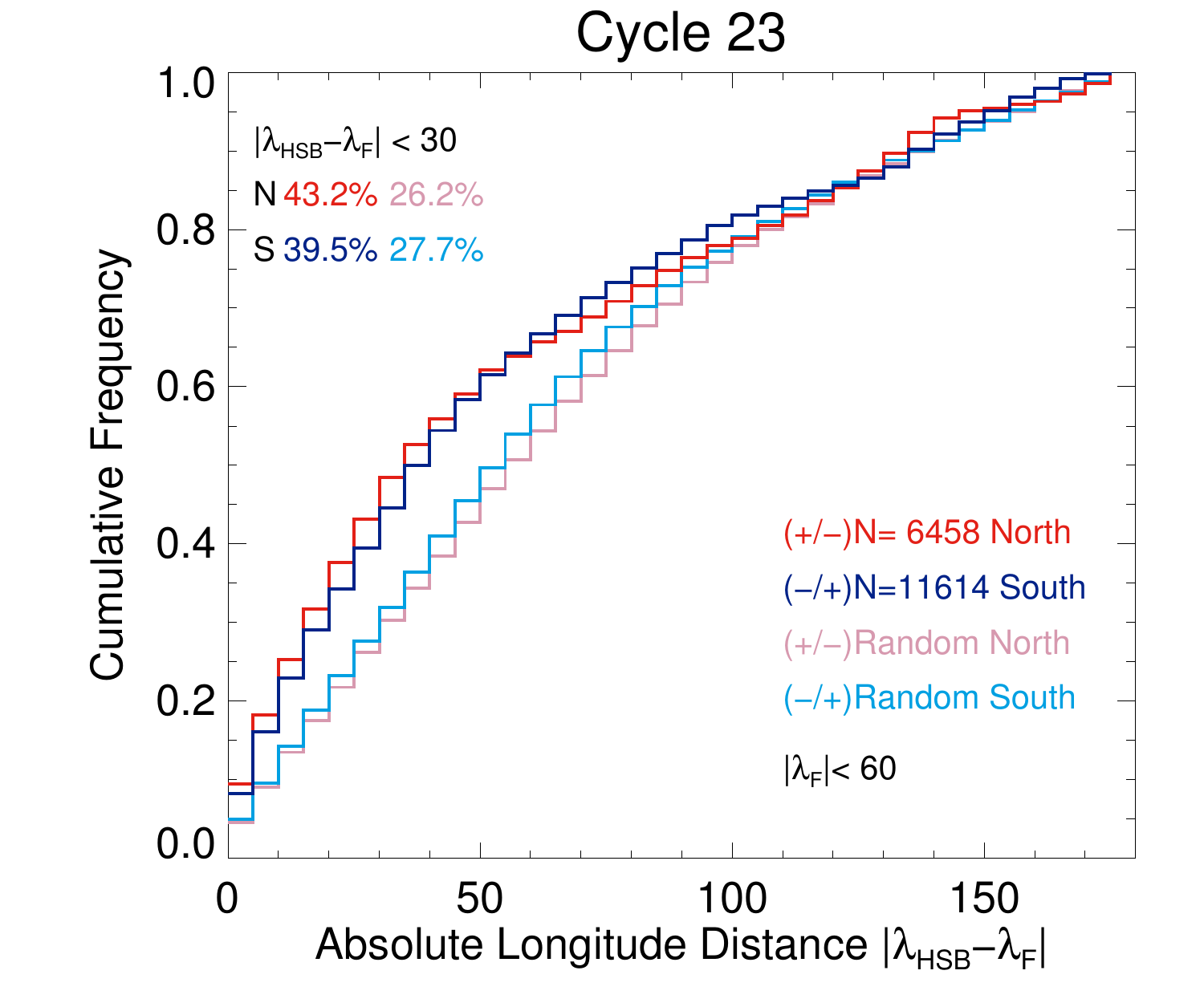}
\includegraphics[width=0.4\textwidth]{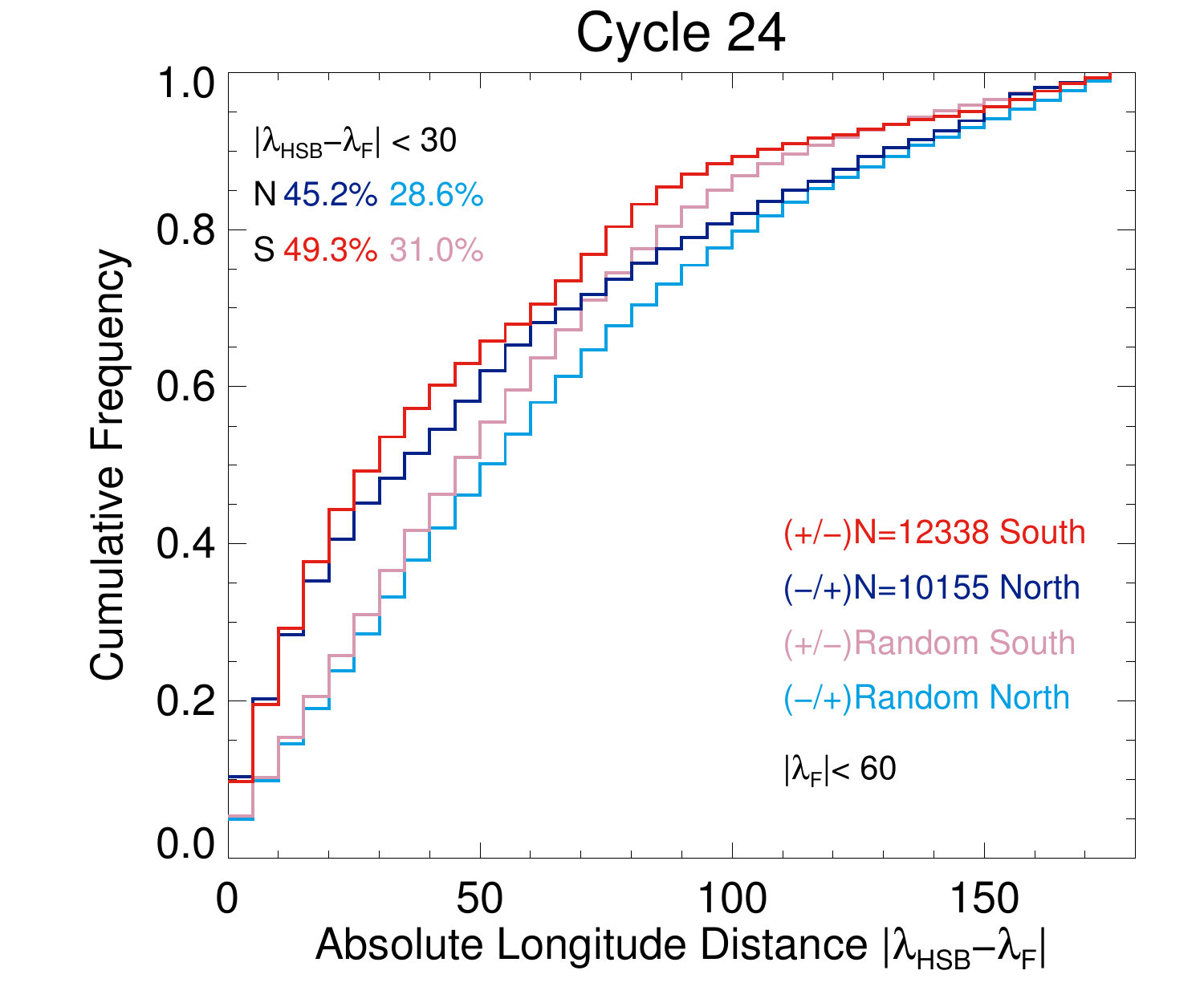}
\caption{Longitudinal distance histograms (top row) of RHESSI flares with $|\lambda_{F}|\leqslant60^{\circ}$ from the HSB PFSS, for solar Cycle 23 and 24 (left and right columns). With red, the change in polarity from positive to negative and blue from negative to positive. Cumulative distributions of these histograms (bottom row), with additional cumulative distributions over-plotted that were produced from a set of flares with longitudes drawn from a random distribution (pink for change in polarity from positive to negative and light blue for change in polarity from negative to positive). }  
\label{fig:lesslng}
\end{figure*}

\begin{figure*}\centering  
\includegraphics[width=0.4\textwidth]{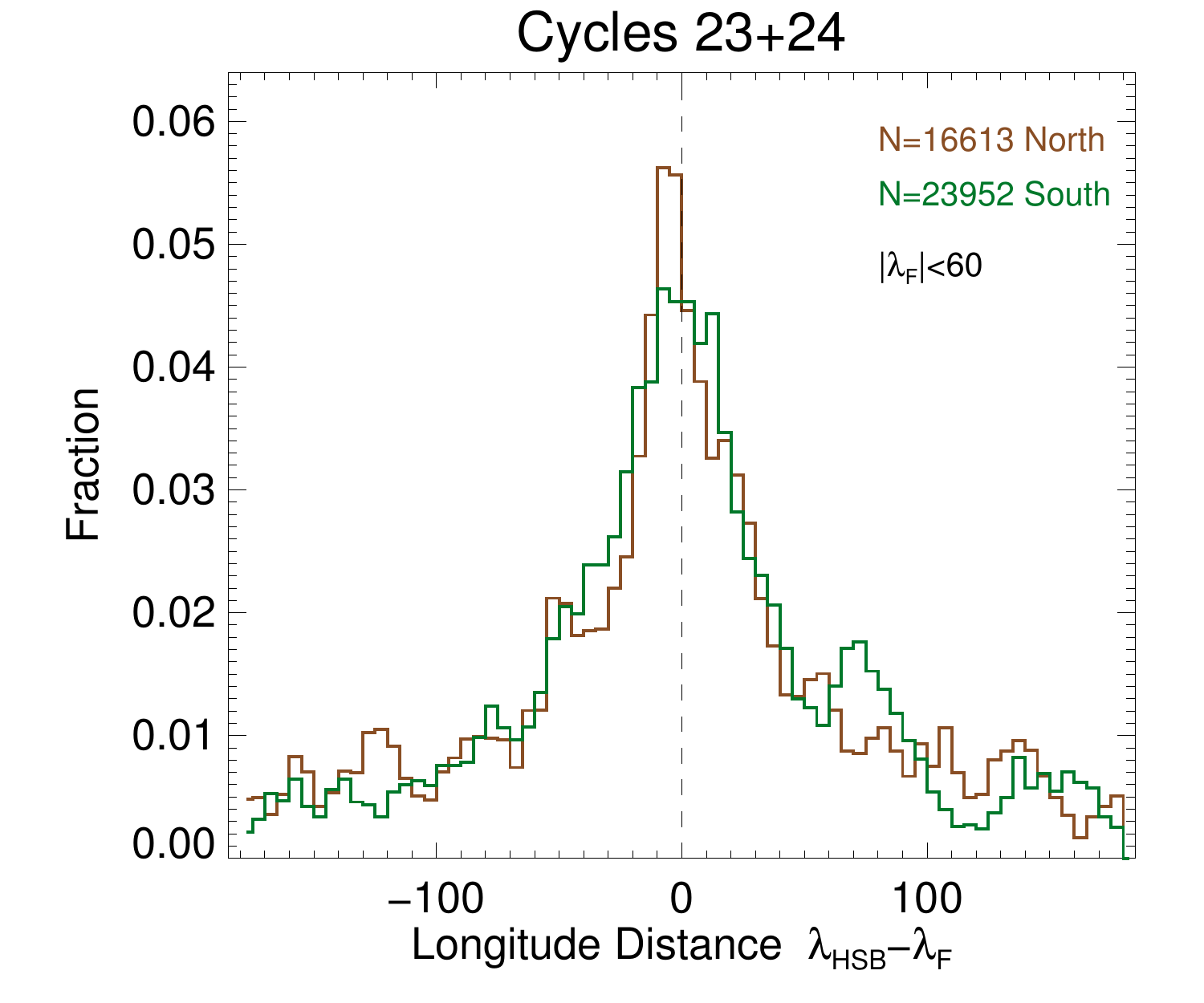}
\includegraphics[width=0.4\textwidth]{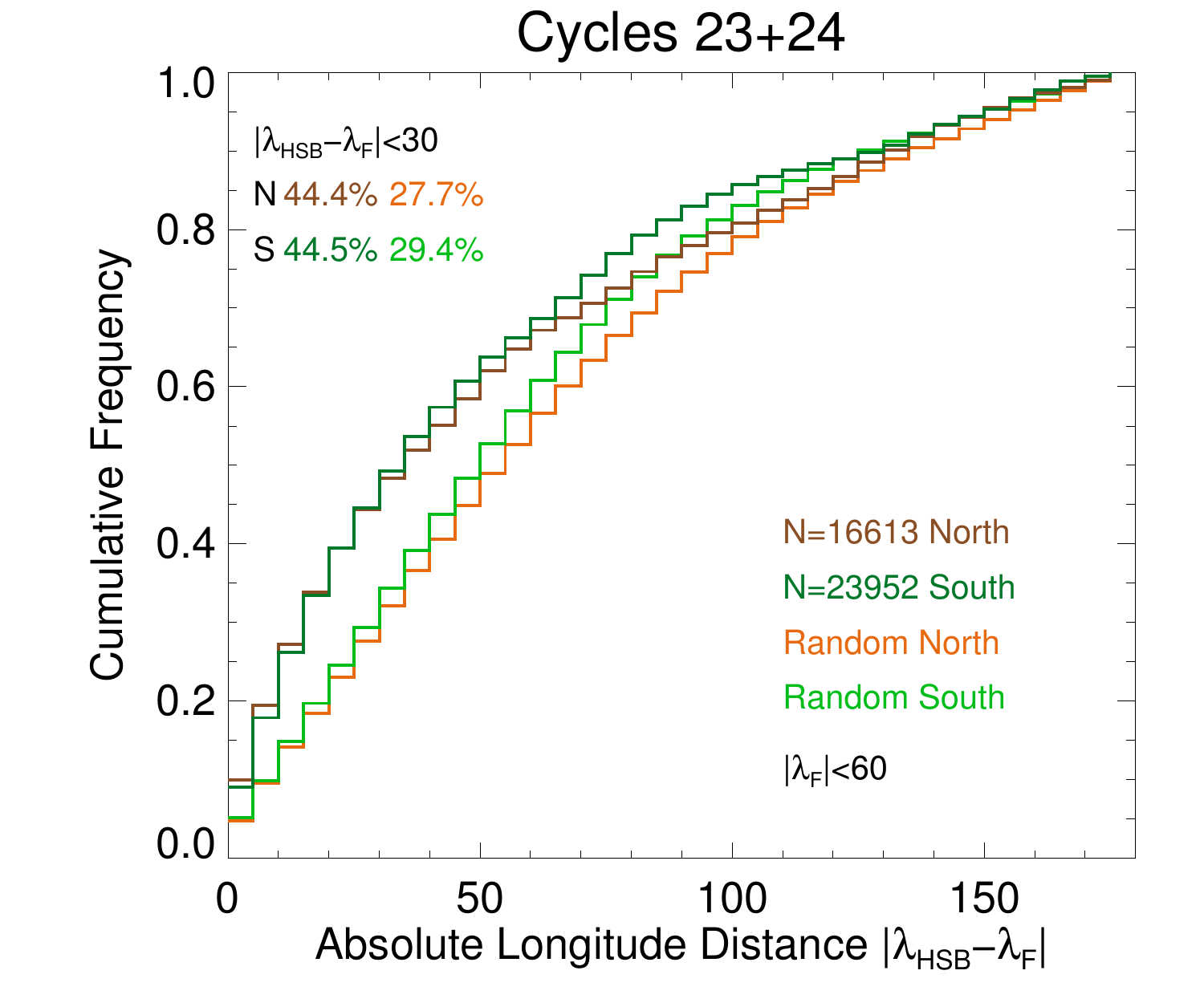}
\caption{Longitudinal distance histograms (left) of RHESSI flares with $|\lambda_{F}|\leqslant60^{\circ}$ from the HSB PFSS, for solar Cycle 23 and 24 combined. With brown, all the flares located in the northern hemisphere and with dark green all the southern flares. Cumulative distributions of these histograms (right), with additional cumulative distributions over-plotted that were produced from a set of flares with longitudes drawn from a random distribution (orange for northern and light green for southern).}  
\label{fig:bothcyc}
\end{figure*}

For each of the RHESSI flares on our restricted list (see Appendix~\ref{nsec:rhessifl}) we calculate the nearest HSB PFSS that occurred on that day at the flare's latitude. Here we add an additional restriction of only using flares away from the limbs $|\lambda_{F}|\leqslant60^{\circ}$, as the position of limb flares have a greater uncertainty due to projection effects. This flare sample therefore differs from that previously in Section~\ref{nsec:findhsbesc}. For Cycle 23 we were able to find the longitude distance to the HSB for 18,072 flares and 22,493 for Cycle 24. The top row of Fig. \ref{fig:lesslng} shows the resulting histograms of the longitude distance between the RHESSI flares and HSB PFSS again separated by Cycle and polarity change/hemisphere. For both Cycles we see a peak for small distances between the flare location and HSB. For Cycle 23, the northern hemisphere is more sharply peaked than the southern, with a number of southern flare longitudes further round than the HSB ($\lambda_{HSB}-\lambda_{F}<0$). In Cycle 24 the distributions peak in a similar manner, though again it is slightly flatter for southern flares near the HSB ($\lambda_{HSB}-\lambda_{F}\simeq0$). There is also a number  of southern flares with $\lambda_{HSB}-\lambda_{F}\simeq70^{\circ}$. At the bottom row of Fig. \ref{fig:lesslng} we also show the cumulative distributions of the flare-HSB longitude distances. To verify that our results are statistically significant, we repeat the analysis, but using flare longitudes randomly drawn from a uniform distribution. The resulting cumulative distributions for the real and random flare locations are shown in Fig. \ref{fig:lesslng} (bottom row). For Cycle 23, we find about 43\% of northern hemisphere flares, and 40\% of the southern occur within $|\lambda_{HSB}-\lambda_{F}|\leq 30^\circ$. For Cycle 24, 45\% of northern hemisphere flares, and  49\% of the southern occur within $|\lambda_{HSB}-\lambda_{F}|\leq 30^\circ$. As expected, the random sample gives a much weaker association (Cycle 23, 26\% north, 28\% south; Cycle 24, 29\% north, 31\% south) showing that the flare-HSB association is statistically significant.

 We finally combine the flare-HSB longitude distance over both Cycles, but still separated by hemisphere, with the resulting distributions shown in Fig. \ref{fig:bothcyc}. Both hemispheres show similar spreads in the distribution, but again the southern hemisphere is slightly less peaked about $\lambda_{HSB}-\lambda_{F}\simeq0$ than the northern. Again the flare-HSB association is considerably weaker for the random sample of flare longitudes. Overall, we have about 44\% of flares occurring within $|\lambda_{HSB}-\lambda_{F}|\leq 30^\circ$.

\section{Discussion \& Conclusions}\label{nsec:discon}

With both the HSB-Earth and HSB PFSS approaches we have been able to show that this magnetic phenomenon maps back to the solar surface and is associated with RHESSI X-ray flares.This HSB-flare association is shown quantitatively in Figs.~\ref{fig:lesslng} and \ref{fig:bothcyc}, finding that overall nearly half of RHESSI flares occur within a longitude of $\pm30^\circ$ of a HSB. Our extension of the work of \citet{2011ApJ...733...49S}, has shown that the HSB hemispheres do swap as we change from Cycle 23 to 24. However the HSB-Earth association is not as clear in Cycle 24, especially in the northern hemisphere. Our comparison between this and the HSB PFSS approach seems to suggest that in Cycle 24 the mean solar wind is faster than we assumed, with the actual HSB lagging behind where HSB-Earth predicted it to be. \citet{2017SoPh..292..174G} used the observed solar wind speed to determine the time of HSB at central meridian, finding a sharper association to flares. Although even this approach found a weaker occurrence pattern for Cycle 24 in both hemispheres compared to Cycle 23. Despite this, we are still able to show a concentration of flares in the expected hemisphere associated with the HSB, even separated by GOES class. The same association works for large X-class flares down to A-class microflares. 

Determining the HSB via PFSS allows it to be found for all times and flares. This provides a clearer picture of the sector structure closer to the solar surface, and one that is not affected by variations in the solar wind speed. From this approach we find that about $41\%$ (Cycle 23),  $47\%$ (Cycle 24) of RHESSI flares occur within $30^\circ$ of the HSB PFSS. The advantage of the PFSS approach is that we achieve a near-instantaneous view of the current structures on the Sun. We  therefore provide a webpage\footnote{\url{http://www.astro.gla.ac.uk/~iain/hsb}} where the daily HSB PFSS are shown relative to the current active regions and recent flare activity.

The confirmation that the HSB and active longitudes do overlap sometimes is expected, since the former has an association with activity and the latter is directly derived from activity indicators. However it is clear that they are not exactly the same thing, and show differences in migration paths; the parabolic passage of the active longitudes over time, as found by \citep{2005A&A...441..347U, 2016ApJ...818..127G} are not apparent in the HSB. The closest matching occurs when both are rotating at about, or slightly faster, than the Carrington rate, with the ``flip-flop'' between dominate active longitudes seen as jumping between a two-sector structure in that hemisphere (during a period of overall four-sectors). Considering the difference in Carrington phase between the active longitudes and HSB does help quantify this association, with some concentration near zero. However this varies between hemispheres and the majority of phase differences are positive, showing that the HSB Carrington phase is mostly larger than the active longitude's. This might be due to the observed migration paths, with the HSB mostly appearing to steadily increase in phase with time, whereas the active longitudes' phases increase and decrease with time. However there are several limitations and assumptions to this HSB and active longitude comparison presented in this paper that merit investigation in future studies. It is problematic that the active longitudes are not detected at all times, and linear interpolation has to be used for a full comparison to the HSB. In addition, we only considered the most dominant active longitudes, so future work could benefit from comparing the HSB to both active longitude bands. Using the HSB found from the PFSS approach could help, removing uncertainties arising from a variable solar wind speed, but this would be limited to more recent Cycles due to the availability of suitable magnetograms. Another discrepancy is that the identified active longitudes persist across several cycles while the HSB of each type swaps hemisphere as the cycle changes. As this occurs during solar minimum, it is not a major issue for the HSB-flare association, but would certainly merit further work to determine how the location of the HSB evolves as the Cycles changes, especially in comparison to the active longitudes.

It is a remarkable feature that a magnetic phenomenon detected at the Earth can be mapped back to activity on the Sun, providing a fast and simple approach to determine longitudinal regions of concentrated activity compared to the longterm averaging and filtering required in active longitude studies. Given that the HSB can often be determined at times when the active longitude cannot be found, a combination of both could be vital for not only helping to predict future activity but for understanding the internal source of the phenomena. It is possible to speculate as to why active emerges repeatedly at similar longitudinal regions \citep[e.g.][]{2004ApJ...604..944B,2006A&A...445..703B,2007AdSpR..40..951U}, but why this should be reflected in the large-scale magnetic field through the HSB is still not clear and requires further investigation.


\begin{acknowledgements}

KL and IGH are funded via a University of Glasgow Lord Kelvin-Adam Smith Fellowship. IGH is supported by a Royal Society University Research Fellowship.  The authors thank the International Space Science Institute (ISSI) for support for the team ``Improving the Reliability of Solar Eruption Predictions to Facilitate the Determination of Targets-of-Opportunity for Instruments With a Limited Field-of-View'', where this work benefited from productive discussions. The authors would like to thank N. Gyenge for very helpful discussions about active longitudes and providing the data from \citet{2016ApJ...818..127G}.

\end{acknowledgements}


\begin{thebibliography}{43}
\expandafter\ifx\csname natexlab\endcsname\relax\def\natexlab#1{#1}\fi

\bibitem[{{Akasofu}(2015)}]{2015GeoRL..42.2571A}
{Akasofu}, S.-I. 2015, \grl, 42, 2571

\bibitem[{{Altschuler} \& {Newkirk}(1969)}]{1969SoPh....9..131A}
{Altschuler}, M.~D. \& {Newkirk}, G. 1969, \solphys, 9, 131

\bibitem[{{Antonucci} \& {Svalgaard}(1974)}]{1974SoPh...36..115A}
{Antonucci}, E. \& {Svalgaard}, L. 1974, \solphys, 36, 115

\bibitem[{{Bai}(2003)}]{2003ApJ...585.1114B}
{Bai}, T. 2003, \apj, 585, 1114

\bibitem[{{Baranyi} {et~al.}(2016){Baranyi}, {Gy{\H o}ri}, \&
  {Ludm{\'a}ny}}]{2016SoPh..291.3081B}
{Baranyi}, T., {Gy{\H o}ri}, L., \& {Ludm{\'a}ny}, A. 2016, \solphys, 291, 3081

\bibitem[{{Berdyugina} {et~al.}(2006){Berdyugina}, {Moss}, {Sokoloff}, \&
  {Usoskin}}]{2006A&A...445..703B}
{Berdyugina}, S.~V., {Moss}, D., {Sokoloff}, D., \& {Usoskin}, I.~G. 2006,
  \aap, 445, 703

\bibitem[{{Berdyugina} \& {Usoskin}(2003)}]{2003A&A...405.1121B}
{Berdyugina}, S.~V. \& {Usoskin}, I.~G. 2003, \aap, 405, 1121

\bibitem[{{Bigazzi} \& {Ruzmaikin}(2004)}]{2004ApJ...604..944B}
{Bigazzi}, A. \& {Ruzmaikin}, A. 2004, \apj, 604, 944

\bibitem[{{Bumba} \& {Obridko}(1969)}]{1969SoPh....6..104B}
{Bumba}, V. \& {Obridko}, V.~N. 1969, \solphys, 6, 104

\bibitem[{{Carrington}(1863)}]{1863spots...C}
{Carrington}, R.~C. 1863, {Observations of the Spots on the Sun from November
  9, 1853 to March 24, 1861, Made at Redhill} (Williams and Norgate)

\bibitem[{{Christe} {et~al.}(2008){Christe}, {Hannah}, {Krucker}, {McTiernan},
  \& {Lin}}]{2008ApJ...677.1385C}
{Christe}, S., {Hannah}, I.~G., {Krucker}, S., {McTiernan}, J., \& {Lin}, R.~P.
  2008, \apj, 677, 1385

\bibitem[{{Crawford} {et~al.}(1970){Crawford}, {Jauncey}, \&
  {Murdoch}}]{1970ApJ...162..405C}
{Crawford}, D.~F., {Jauncey}, D.~L., \& {Murdoch}, H.~S. 1970, \apj, 162, 405

\bibitem[{{Dittmer}(1975)}]{1975SoPh...41..227D}
{Dittmer}, P.~H. 1975, \solphys, 41, 227

\bibitem[{{Gaizauskas} {et~al.}(1983){Gaizauskas}, {Harvey}, {Harvey}, \&
  {Zwaan}}]{1983ApJ...265.1056G}
{Gaizauskas}, V., {Harvey}, K.~L., {Harvey}, J.~W., \& {Zwaan}, C. 1983, \apj,
  265, 1056

\bibitem[{{Getachew} {et~al.}(2017){Getachew}, {Virtanen}, \&
  {Mursula}}]{2017SoPh..292..174G}
{Getachew}, T., {Virtanen}, I., \& {Mursula}, K. 2017, \solphys, 292, 174

\bibitem[{{Gyenge} {et~al.}(2012){Gyenge}, {Baranyi}, \&
  {Ludm{\'a}ny}}]{2012CEAB...36....9G}
{Gyenge}, N., {Baranyi}, T., \& {Ludm{\'a}ny}, A. 2012, Central European
  Astrophysical Bulletin, 36, 9

\bibitem[{{Gyenge} {et~al.}(2014){Gyenge}, {Baranyi}, \&
  {Ludm{\'a}ny}}]{2014SoPh..289..579G}
{Gyenge}, N., {Baranyi}, T., \& {Ludm{\'a}ny}, A. 2014, \solphys, 289, 579

\bibitem[{{Gyenge} {et~al.}(2016){Gyenge}, {Ludm{\'a}ny}, \&
  {Baranyi}}]{2016ApJ...818..127G}
{Gyenge}, N., {Ludm{\'a}ny}, A., \& {Baranyi}, T. 2016, \apj, 818, 127

\bibitem[{{Gyenge} {et~al.}(2017){Gyenge}, {Singh}, {Kiss}, {Srivastava}, \&
  {Erd{\'e}lyi}}]{2017ApJ...838...18G}
{Gyenge}, N., {Singh}, T., {Kiss}, T.~S., {Srivastava}, A.~K., \&
  {Erd{\'e}lyi}, R. 2017, \apj, 838, 18

\bibitem[{{Gy{\H o}ri} {et~al.}(2017){Gy{\H o}ri}, {Ludm{\'a}ny}, \&
  {Baranyi}}]{2017MNRAS.465.1259G}
{Gy{\H o}ri}, L., {Ludm{\'a}ny}, A., \& {Baranyi}, T. 2017, \mnras, 465, 1259

\bibitem[{{Hale} {et~al.}(1919){Hale}, {Ellerman}, {Nicholson}, \&
  {Joy}}]{1919ApJ....49..153H}
{Hale}, G.~E., {Ellerman}, F., {Nicholson}, S.~B., \& {Joy}, A.~H. 1919, \apj,
  49, 153

\bibitem[{{Hannah} {et~al.}(2011){Hannah}, {Hudson}, {Battaglia}, {Christe},
  {Ka{\v s}parov{\'a}}, {Krucker}, {Kundu}, \& {Veronig}}]{2011SSRv..159..263H}
{Hannah}, I.~G., {Hudson}, H.~S., {Battaglia}, M., {et~al.} 2011, \ssr, 159,
  263

\bibitem[{{Hudson}(1991)}]{1991SoPh..133..357H}
{Hudson}, H.~S. 1991, \solphys, 133, 357

\bibitem[{{Hudson} {et~al.}(2014){Hudson}, {Svalgaard}, \&
  {Hannah}}]{2014SSRv..186...17H}
{Hudson}, H.~S., {Svalgaard}, L., \& {Hannah}, I.~G. 2014, \ssr, 186, 17

\bibitem[{{Hurford} {et~al.}(2002){Hurford}, {Schmahl}, {Schwartz}, {Conway},
  {Aschwanden}, {Csillaghy}, {Dennis}, {Johns-Krull}, {Krucker}, {Lin},
  {McTiernan}, {Metcalf}, {Sato}, \& {Smith}}]{2002SoPh..210...61H}
{Hurford}, G.~J., {Schmahl}, E.~J., {Schwartz}, R.~A., {et~al.} 2002, \solphys,
  210, 61

\bibitem[{{Li}(2011)}]{2011ApJ...735..130L}
{Li}, J. 2011, \apj, 735, 130

\bibitem[{{Lin} {et~al.}(2002){Lin}, {Dennis}, {Hurford}, {Smith}, {Zehnder},
  {Harvey}, {Curtis}, {Pankow}, {Turin}, {Bester}, {Csillaghy}, {Lewis},
  {Madden}, {van Beek}, {Appleby}, {Raudorf}, {McTiernan}, {Ramaty}, {Schmahl},
  {Schwartz}, {Krucker}, {Abiad}, {Quinn}, {Berg}, {Hashii}, {Sterling},
  {Jackson}, {Pratt}, {Campbell}, {Malone}, {Landis}, {Barrington-Leigh},
  {Slassi-Sennou}, {Cork}, {Clark}, {Amato}, {Orwig}, {Boyle}, {Banks},
  {Shirey}, {Tolbert}, {Zarro}, {Snow}, {Thomsen}, {Henneck}, {McHedlishvili},
  {Ming}, {Fivian}, {Jordan}, {Wanner}, {Crubb}, {Preble}, {Matranga}, {Benz},
  {Hudson}, {Canfield}, {Holman}, {Crannell}, {Kosugi}, {Emslie}, {Vilmer},
  {Brown}, {Johns-Krull}, {Aschwanden}, {Metcalf}, \&
  {Conway}}]{2002SoPh..210....3L}
{Lin}, R.~P., {Dennis}, B.~R., {Hurford}, G.~J., {et~al.} 2002, \solphys, 210,
  3

\bibitem[{{Losh}(1939)}]{1939POMic...7..127L}
{Losh}, H.~M. 1939, Publications of Michigan Observatory, 7, 127

\bibitem[{{Maunder}(1904)}]{1904MNRAS..64..747M}
{Maunder}, E.~W. 1904, \mnras, 64, 747

\bibitem[{{Owens} \& {Forsyth}(2013)}]{2013LRSP...10....5O}
{Owens}, M.~J. \& {Forsyth}, R.~J. 2013, Living Reviews in Solar Physics, 10, 5

\bibitem[{{Pelt} {et~al.}(2010){Pelt}, {Korpi}, \&
  {Tuominen}}]{2010A&A...513A..48P}
{Pelt}, J., {Korpi}, M.~J., \& {Tuominen}, I. 2010, \aap, 513, A48

\bibitem[{{Rosenberg} \& {Coleman}(1969)}]{1969JGR....74.5611R}
{Rosenberg}, R.~L. \& {Coleman}, Jr., P.~J. 1969, \jgr, 74, 5611

\bibitem[{{Schatten} {et~al.}(1969){Schatten}, {Wilcox}, \&
  {Ness}}]{1969SoPh....6..442S}
{Schatten}, K.~H., {Wilcox}, J.~M., \& {Ness}, N.~F. 1969, \solphys, 6, 442

\bibitem[{{Schulz}(1973)}]{1973Ap&SS..24..371S}
{Schulz}, M. 1973, \apss, 24, 371

\bibitem[{{Schwartz} {et~al.}(2002){Schwartz}, {Csillaghy}, {Tolbert},
  {Hurford}, {McTiernan}, \& {Zarro}}]{2002SoPh..210..165S}
{Schwartz}, R.~A., {Csillaghy}, A., {Tolbert}, A.~K., {et~al.} 2002, \solphys,
  210, 165

\bibitem[{{Svalgaard} {et~al.}(2011){Svalgaard}, {Hannah}, \&
  {Hudson}}]{2011ApJ...733...49S}
{Svalgaard}, L., {Hannah}, I.~G., \& {Hudson}, H.~S. 2011, \apj, 733, 49

\bibitem[{{Svalgaard} \& {Wilcox}(1976)}]{1976SoPh...49..177S}
{Svalgaard}, L. \& {Wilcox}, J.~M. 1976, \solphys, 49, 177

\bibitem[{{Usoskin} {et~al.}(2007){Usoskin}, {Berdyugina}, {Moss}, \&
  {Sokoloff}}]{2007AdSpR..40..951U}
{Usoskin}, I.~G., {Berdyugina}, S.~V., {Moss}, D., \& {Sokoloff}, D.~D. 2007,
  Advances in Space Research, 40, 951

\bibitem[{{Usoskin} {et~al.}(2005){Usoskin}, {Berdyugina}, \&
  {Poutanen}}]{2005A&A...441..347U}
{Usoskin}, I.~G., {Berdyugina}, S.~V., \& {Poutanen}, J. 2005, \aap, 441, 347

\bibitem[{{Wheatland}(2004)}]{2004ApJ...609.1134W}
{Wheatland}, M.~S. 2004, \apj, 609, 1134

\bibitem[{{Wilcox} \& {Ness}(1965)}]{1965JGR....70.5793W}
{Wilcox}, J.~M. \& {Ness}, N.~F. 1965, \jgr, 70, 5793

\bibitem[{{Zhang} {et~al.}(2011){Zhang}, {Mursula}, {Usoskin}, \&
  {Wang}}]{2011JASTP..73..258Z}
{Zhang}, L., {Mursula}, K., {Usoskin}, I., \& {Wang}, H. 2011, Journal of
  Atmospheric and Solar-Terrestrial Physics, 73, 258

\bibitem[{{Zwaan}(1987)}]{1987ARA&A..25...83Z}
{Zwaan}, C. 1987, \araa, 25, 83

\end{thebibliography}


\begin{appendix}

\section{The RHESSI flare list}\label{nsec:rhessifl}

The flares for this study were found using the X-ray observations of the Reuven Ramaty High Energy Solar Spectroscopic Imager RHESSI \citep{2002SoPh..210....3L}. RHESSI started observing in February 2002 (about halfway through Cycle 23) and continues to take data through Cycle 24. RHESSI has several advantages for flares studies over other instruments: it observes at higher energies where the flare to solar background contrast is higher; its attenuating shutters allow it to observe from the largest X-class flares down to A-class microflares. Hard X-ray flare durations are shorter and so there is less confusion for weaker events. Significantly, RHESSI also provides spatial information about the flares, which is time encoded in its light curves via the Rotation Modulation Collimators in front of the detectors \citep{2002SoPh..210...61H}. This indirect imaging technique means that RHESSI cannot determine the position of flares near its rotation axis, however this is a minimal loss, affecting only about $1.7\%$ events \citep{2008ApJ...677.1385C}. A larger loss of spatial information comes from the Roll Aspect System (RAS) on RHESSI which can produce erroneous flare locations. These have the correct distance from Sun center but at the wrong roll angle. These flares are often easily identified as their position is obviously away from the active regions, often near the limb at high latitudes.

A flare list of RHESSI events is automatically generated from the quicklook data products, by detecting an excess in the 6-12 keV count rate $3\sigma$ above a background level obtained from a 60~second running average. The list and full information about how it is determined is available in the RHESSI branch of Solarsoft, as well as online\footnote{\url{http://sprg.ssl.berkeley.edu/~jimm/hessi/hsi_flare_list.html}} \citep{2002SoPh..210..165S}. For each event many flare characteristics are determined (start, peak and end times, flare location, fluxes etc) as well as flags indicating the state and quality of the RHESSI detection. The latter is important because of the presence of false events within the list. These can arise from the discontinuous nature of the RHESSI data (as solar data are not taken during eclipse or South Atlantic Anomaly  (SAA) passage). In addition non-solar events (e.g. cosmic ray, energetic particles) can be detected as flares but they are straightforward to identify as they give no position.

From 12-Feb-2002 to 23-Feb-2016 there are 117,427 events in the RHESSI flare list\footnote{\url{http://hesperia.gsfc.nasa.gov/hessidata/dbase/hessi_flare_list.txt}}, covering the period from the start of observations to the beginning of RHESSI's fifth detector anneal period (which lasted until 29-April 2016). Then from this full list we produce our restricted list by the following procedure:

\begin{itemize}
  \item We remove flares which the list identifies as having a non-solar position or that have a low front detector to total count rate. We also remove events which were detected with a low data quality, indicating this was during a time of change due to SAA, eclipse or data gaps etc.
  \item The heliographic latitude and longitude of the remaining flares is calculated and we identify those with higher latitudes than expected, $\ge\pm50^\circ$ near the start of the Cycles, down to $\ge\pm20^\circ$ near the end. For each of these events we recalculate the RHESSI flare position using the routine the flare list uses but with a longer time window for the RAS database to try and obtain a more accurate spin rate and hence location. This corrects the position of many flares but some are still left with unusually high latitudes. For these events we manually verify if the RHESSI flare position is associated with an active region or brightening in EUV. In the vast majority of these cases the position cannot be confirmed and so they are dropped from the list.
  \item For each flare we calculate the GOES 1-8\AA~ background subtracted flux by finding the maximum GOES flux between the RHESSI flare start and end time. For flares in which no GOES flux can be determined, or a value is obtained less than A1-class (\begin{math}\mathrm{10^{-8}~Wm^{-2}}\end{math}), we remove them from the list.
\end{itemize}

\begin{figure*}
\centering
 \includegraphics[width=0.8\linewidth]{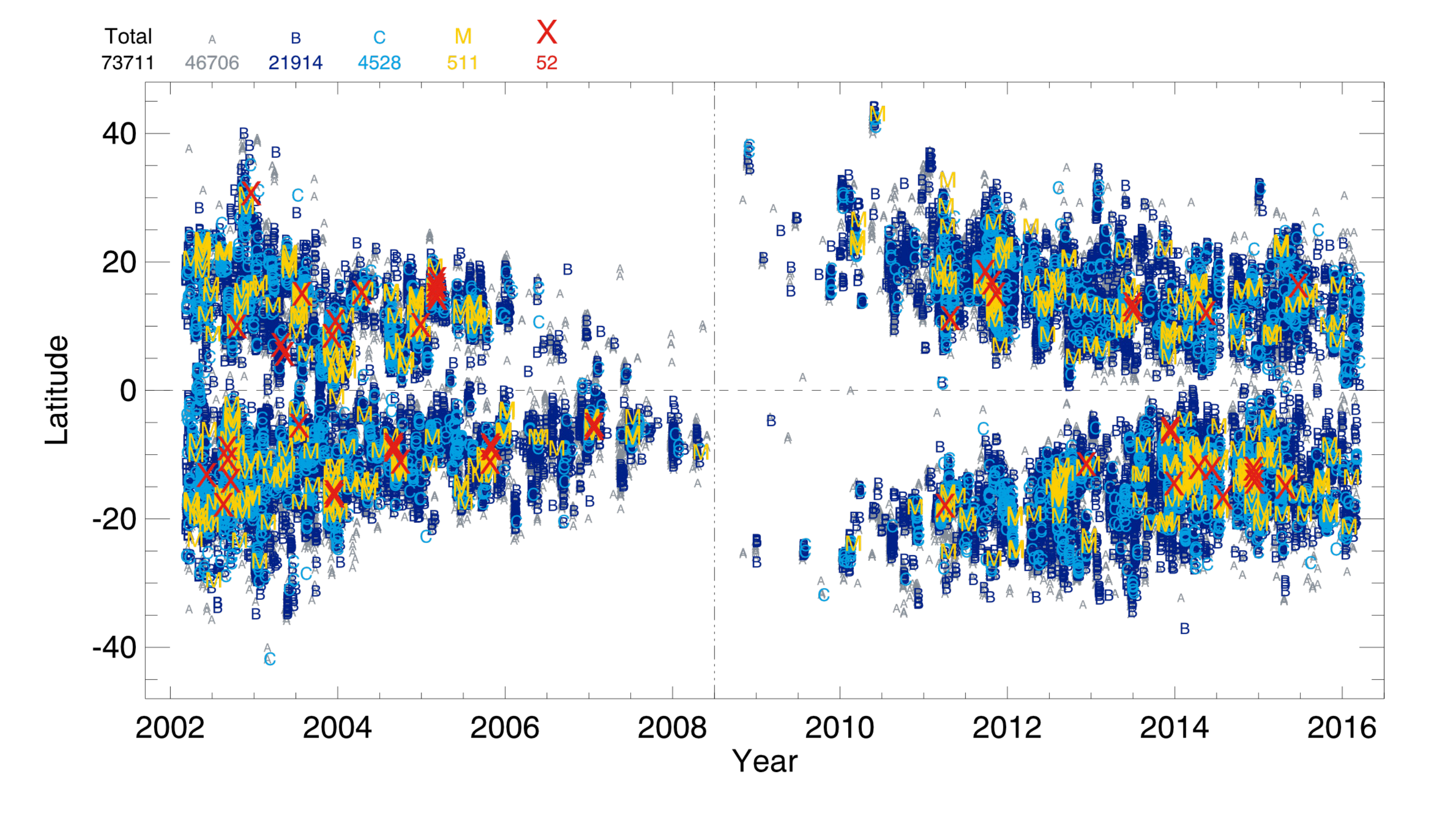}
\caption{The ``Butterfly diagram'' of RHESSI flare latitudes versus time. The colour and letters indicate the different background subtracted GOES class of each flare and the vertical dotted-dash line shows the transition from Cycle 23 to Cycle 24 during the middle of 2008. The vertical gaps (for instance start of 2012 and middle of 2014) are due to extended periods when the RHESSI detector were off due to annealing.} 
\label{fig:butterfly}
\end{figure*}

\begin{figure*}
\centering
 \includegraphics[width=0.3\linewidth]{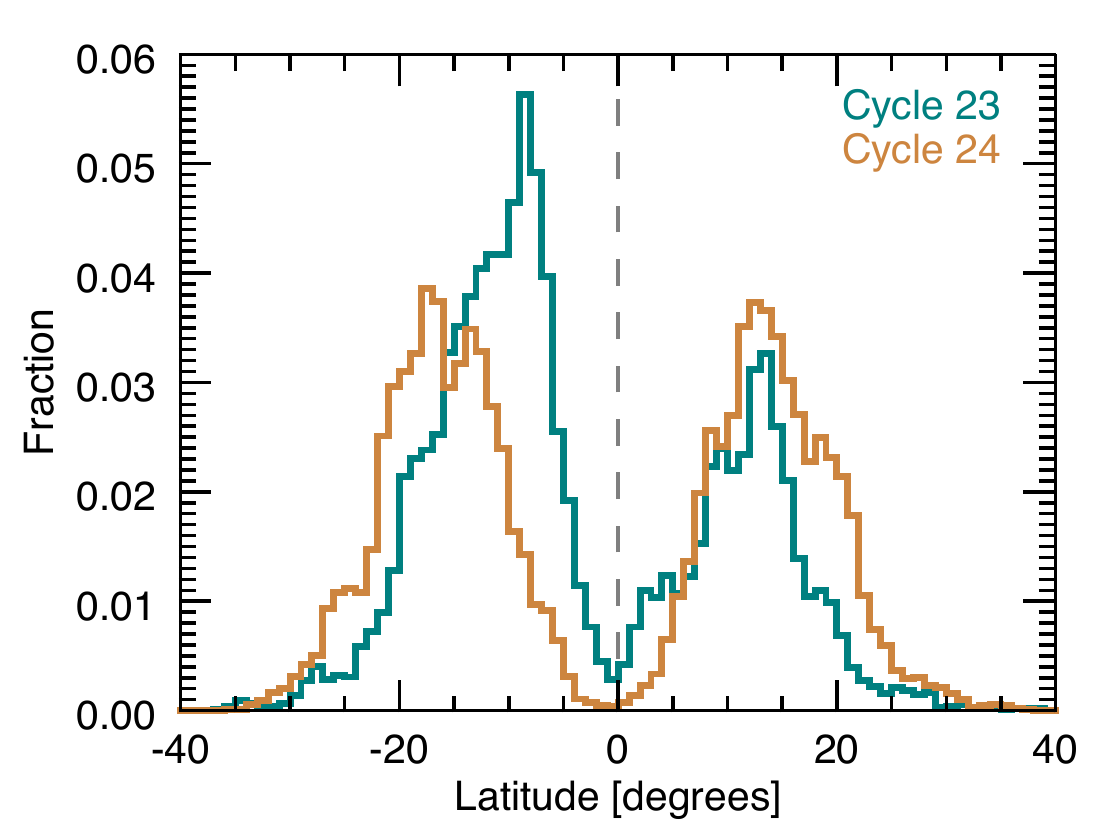}
  \includegraphics[width=0.3\linewidth]{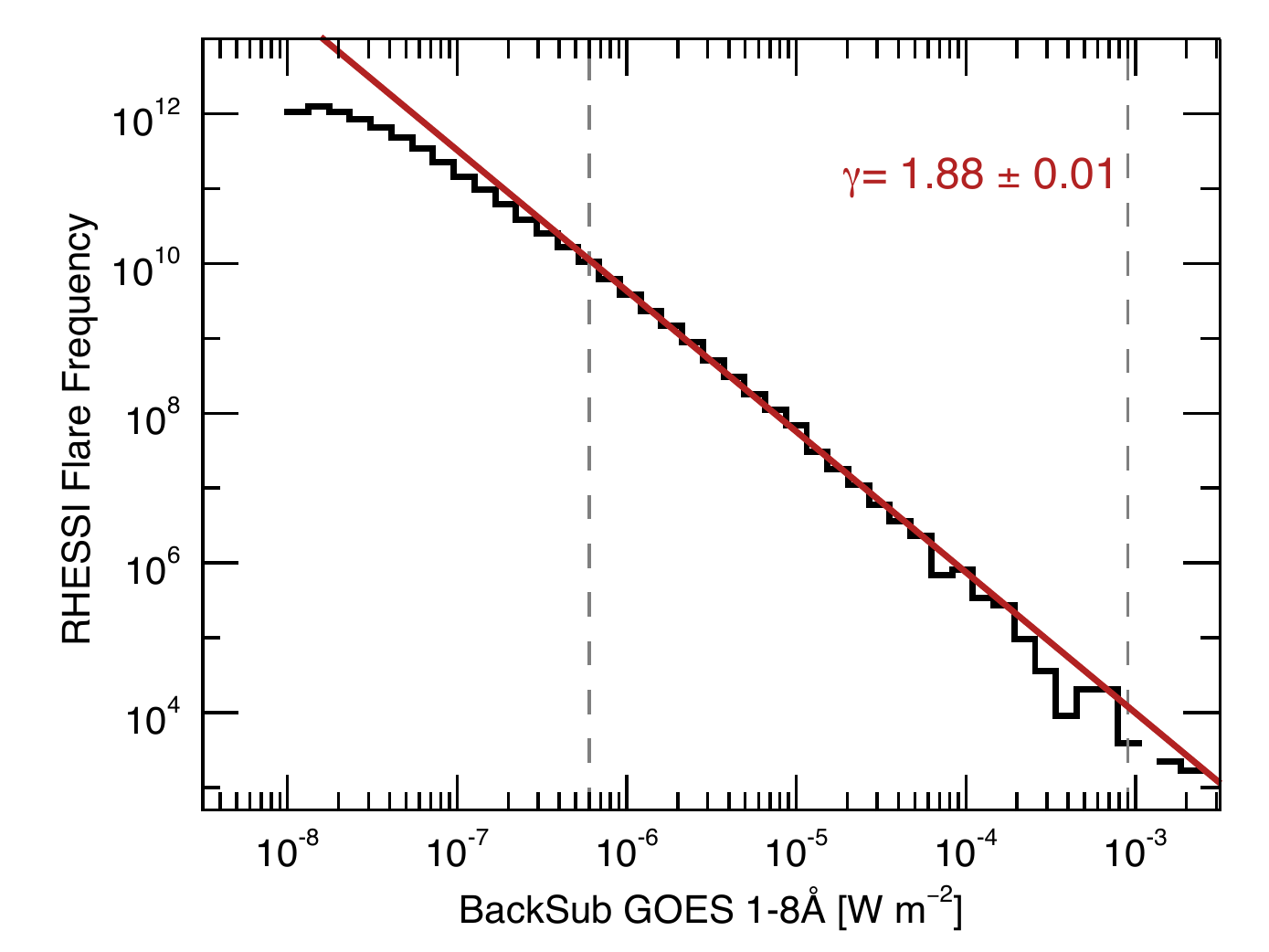}
  \includegraphics[width=0.3\linewidth]{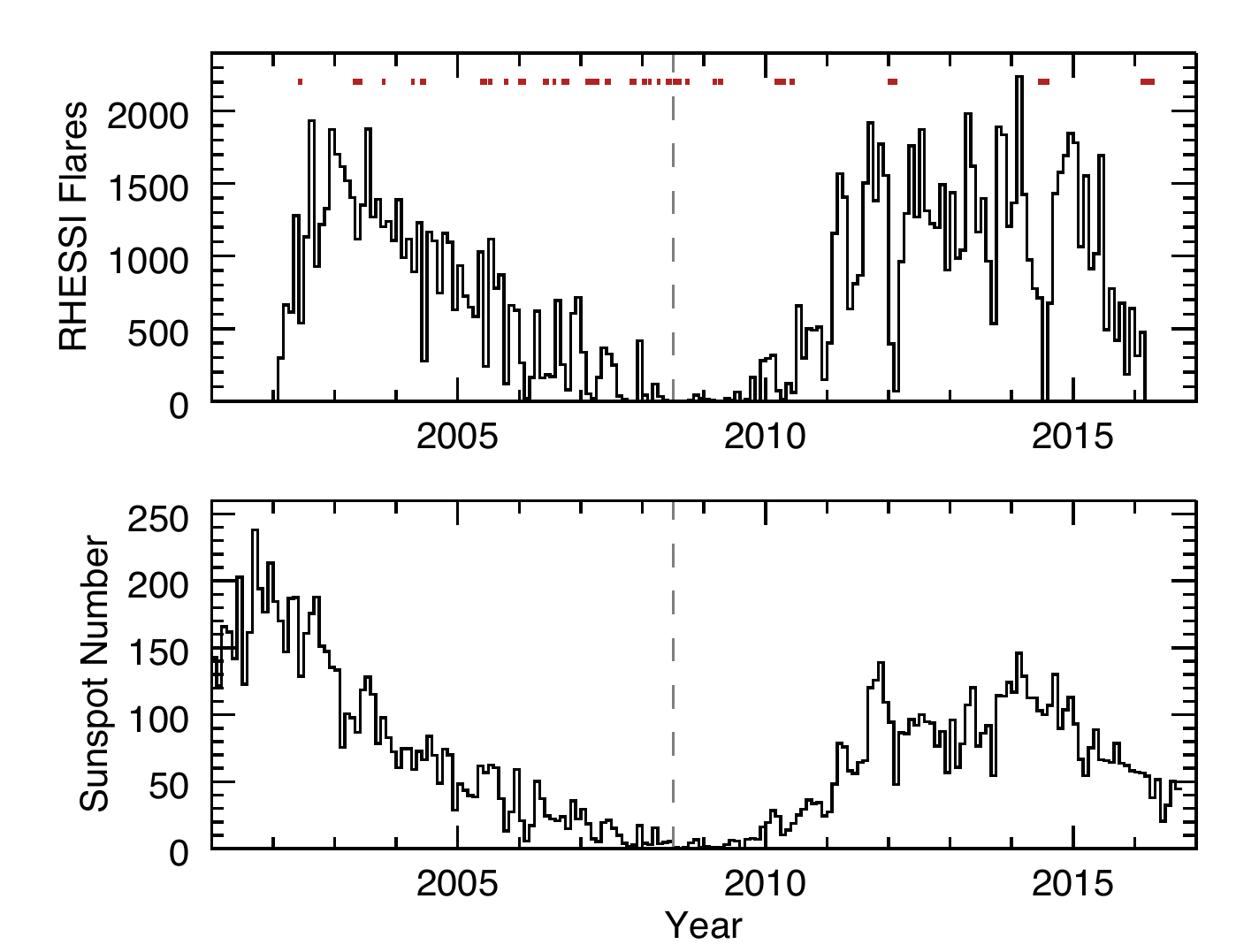}\\
\caption{Summary of the filtered RHESSI flare list showing a histogram of the latitudes per Cycle (left), frequency distribution of the flare GOES 1-8\AA~ flux (middle) and number of flares and sunspots versus time (right). Note that the power-law fitted to the frequency distribution (red line, middle panel) was found using a Maximum Likelihood Estimator method \citep{1970ApJ...162..405C,2004ApJ...609.1134W}, over the range indicated by the dashed vertical lines, and is not a direct fit to the histogram. The times during which RHESSI was not pointing at the Sun or the detectors were off is indicated by the red bars (right panels).} 
\label{fig:listsummary}
\end{figure*}

At the end of the filtering we have 73,711 events. We have tried to minimise the number of dubious events in our final list but there may still be some, particularly from single flares being counted as multiple events due to a disruption in the data. To verify our final list we produce summary plots in Figs. \ref{fig:butterfly} and \ref{fig:listsummary} of the flare characteristics, checking that the broad statistical behaviour of the events is as expected. A Maunder's ``Butterfly diagram'' of the RHESSI flare latitudes in Fig. \ref{fig:butterfly} shows the expected pattern of flares occurring at lower latitudes as the Cycle progresses. From this diagram we also take the transition from Cycle 23 to 24 occurring in mid-2008. A histogram of these latitudes (left, Fig. \ref{fig:listsummary}) shows that there has been a similar distribution of flares between the northern and southern hemisphere during Cycle 24 but a dominance of southern flares in Cycle 23. In Fig. \ref{fig:listsummary} (middle) we also show the flare frequency distribution of the background subtracted GOES flux of each flare. As expected the smaller events are considerably more frequent, with them following a power-law of about -2 \citep{1991SoPh..133..357H}. The distribution of events deviates from the power law for the smallest events, which is expected as they are under sampled due to selection bias, these smaller events being harder to detect, or completely hidden, by the background emission or by larger flares \citep{2011SSRv..159..263H}. 

We also show that the monthly number of RHESSI flares (corrected for the fraction of time lost to eclipse and SAA) approximately matches the number of sunspots (right, Fig. \ref{fig:listsummary}). There are deviations between the two but these can be accounted for due to RHESSI's instrumental performance. There are fewer flares at the start of the mission due to the attenuator being left in, producing a lower sensitivity for smaller, and more frequent, flares. There are several times\footnote{\url{http://sprg.ssl.berkeley.edu/~tohban/offtimes.txt}} during RHESSI's mission when it was either not pointing at the Sun (for observing the Crab Nebula or for quiet Sun studies) or the detectors were off and annealing (heated to recover performance), indicated by a red bars in Fig. \ref{fig:listsummary}. Subsequently, there are no flares in the list during these times. There is also a sharper fall off in the number of flares during 2015 into 2016 than in sunspot number, which is likely due to RHESSI's reduced sensitivity and higher background prior to the 2016 anneal.

\end{appendix}


\end{document}